\documentclass[superscriptaddress,longbibliography,onecolumn, preprint]{revtex4-2}

\usepackage{cancel,soul,ulem}

\usepackage{xcolor}
\usepackage{amsmath,amssymb,graphicx}
\usepackage{graphicx}
\usepackage{epstopdf}
\usepackage{mathtools}
\usepackage{upgreek}
\usepackage{notes2bib}
\usepackage{bm}
\usepackage{color}
\usepackage{braket}
\usepackage{textgreek}
\usepackage{soul}
\usepackage{lipsum}
\usepackage{lineno}
\setstcolor{red}

\begin{document}
\baselineskip24pt

\title{Observation of non-Hermitian topology from optical loss modulation}

\author{Amin Hashemi}
\affiliation{CREOL, The College of Optics and Photonics, University of Central Florida, Orlando, FL 32816, USA}

\author{Elizabeth Louis Pereira}
\affiliation{Department of Applied Physics, Aalto University, 02150 Espoo, Finland}

\author{Hongwei Li}
\affiliation{Nokia Bell Labs, 21 JJ Thomson Avenue, Cambridge, CB3 0FA, UK}

\author{Jose L. Lado}
\affiliation{Department of Applied Physics, Aalto University, 02150 Espoo, Finland}

\author{Andrea Blanco-Redondo}
\email[Corresponding author: ]{andrea.blancoredondo@ucf.edu}
\affiliation{CREOL, The College of Optics and Photonics, University of Central Florida, Orlando, FL 32816, USA}

\begin{abstract}
Understanding the interplay of non-Hermiticity and topology is crucial given the intrinsic openness of most natural and engineered systems and it has important ramifications in topological lasers and sensors. Intense efforts have been devoted to unveiling how non-Hermiticity may impact the most significant features of topological systems, but only recently it has been theoretically proposed that topological features could originate solely from the system's non-Hermiticity in photonic systems. In this work, we experimentally demonstrate the appearance of non-Hermitian topology exclusively from loss modulation in a photonic system that is topologically trivial in the absence of loss. We do this by implementing a non-Hermitian generalization of an Aubry-André-Harper model with purely imaginary potential in a programmable integrated photonics platform, which allows us to investigate different periodic and quasi-periodic configurations of the model. In both cases, we show the emergence of topological edge modes and explore their resilience to different kinds of disorder. Our work highlights loss engineering as a mechanism to generate topological properties.
\end{abstract}

\maketitle

\section*{Introduction}
The last two decades have witnessed outstanding progress in topological photonics \cite{ozawa2019topological,price2022roadmap}. This has mainly been driven by the prospect of robust electromagnetic modes, protected by global topologies of the dispersion relation of photonic materials, and their applications on classical \cite{bandres2018topological,contractor2022scalable,kumar2022topological} and quantum integrated photonic devices \cite{blanco2018topological,mittal2018topological,dai2022topologically}. While most of this progress has occurred in Hermitian (closed) systems, there have also been notable advances in open systems, where non-Hermiticity and topology interplay to  produce unique features \cite{Nasari:23,Yan_NanPhot2023}.
Non-Hermiticity significantly enriches topological phases beyond the traditional Hermitian framework. By introducing a new class of symmetries \cite{Kawabata_PhysRevX2019}, it substantially expands the existing symmetry classifications, thereby enhancing the complexity and diversity of topological phases beyond what is possible in Hermitian systems.
A most notable feature of non-Hermitian topological systems is that they, generally, have complex eigenvalues, which has vast implications, including the appearance of exceptional points \cite{meng2024exceptional} and the emergence of eigenvalue topology \cite{ding2022non}. 
Building on theoretical investigations \cite{LevitovPRL2009,KohmotoPRB2011,ZollerNatPhys2011,schomerus2013topologically, leykam2017edge,resendiz2020topological}, a number of optical experiments have demonstrated topological effects by incorporating gain or loss: from the observation of stationary zero-order topological states \cite{zeuner2015observation} and topological PT symmetric interface states \cite{weimann2017topologically} in non-Hermitian dimer chains to recent demonstrations of gain-related asymmetric long-range couplings in the Haldane model \cite{liu2021gain} and triple topological phase transitions in topological quasicrystals \cite{weidemann2022topological}, all the way through the extensive literature in topological lasers
\cite{bahari2017nonreciprocal,st2017lasing,zhao2018topological,bandres2018topological,contractor2022scalable}.
All demonstrations to date relied on the introduction of loss and/or gain in systems that were already topological in their Hermitian version \cite{zeuner2015observation,weimann2017topologically,st2017lasing,zhao2018topological,bahari2017nonreciprocal,weidemann2022topological} or used gain in an indirect way to engineer asymmetric couplings in Hermitian systems \cite {liu2021gain}.
Recently, it was theoretically proposed that a photonic topological insulating phase can arise, exclusively, from system's non-Hermiticity through gain and loss modulation in an otherwise topologically trivial system \cite{TakataPRL2018}. Subsequent theoretical studies have shown topological transitions and edge states in non-Hermitian AAH models with imaginary commensurate \cite{BofengPRR2023} and incommensurate \cite{JosePPR2024} imaginary potentials. 
Although experimental demonstrations of topological phases arising solely from non-Hermiticity have occurred in electrical circuits \cite{Electrical_PRApp2020}, acoustics \cite{Acoustic_PRB2020, gao2021non}, elastic waves \cite{Elastic_MechSys2022}, and plasmonics \cite{Plasmonic_PRL2023}, an experimental demonstration remained elusive in photonic systems.\\ 
Here, we experimentally demonstrate the appearance of non-Hermitian topology and the existence of edge modes, exclusively, from optical loss modulation. Our experiments realize a non-Hermitian generalization of the AAH model \cite{JosePPR2024} in a reconfigurable integrated photonics platform \cite{MehmetNatComm2024}. We reveal the presence of topological transitions, the emergence of distinct topological phases, and the presence of topologically protected edge states, for both periodic (commensurate) and quasi-periodic (incommensurate) potentials.
Thanks to the reconfigurability afforded by the programmable photonic platform, we systematically investigate the robustness of the system against different types of disorders. Our experimental results, in conjunction with theoretical analysis, illuminate the intricate interplay between disorder, topology, and the emergence of protected modes in photonic non-Hermitian periodic systems and quasicrystals. These findings may have implications in open topological systems and their applications, such as topological lasers and sensors, where gain and loss play a crucial role.

\section*{Results}
\begin{figure*}
    \centering
    \includegraphics[width=7in]{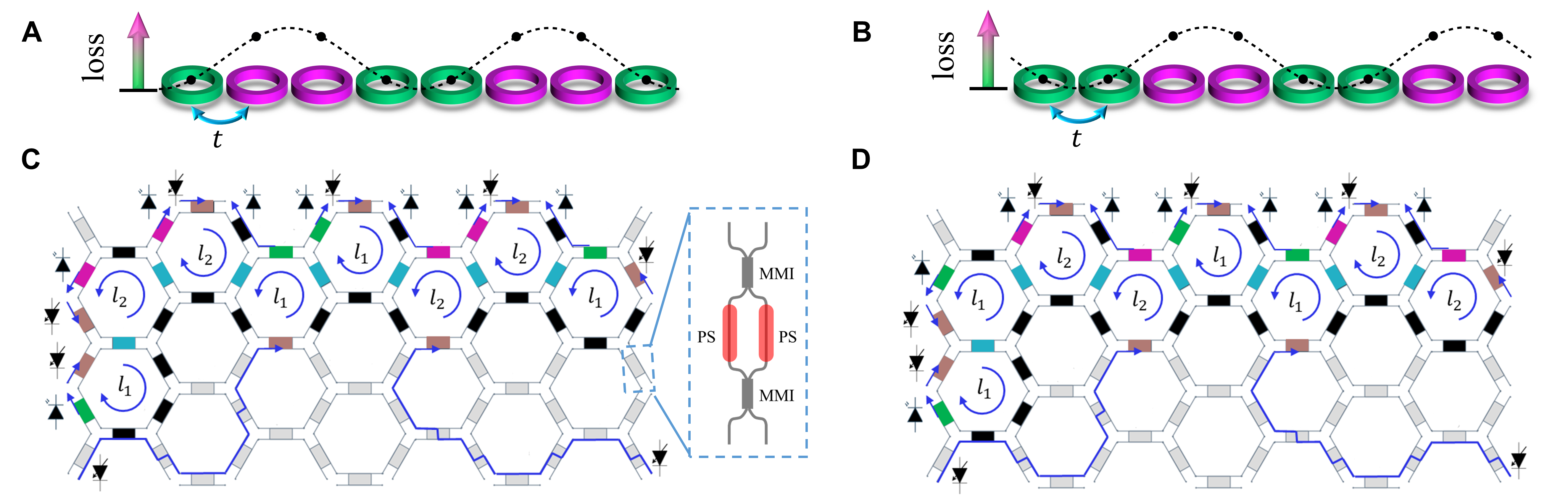}
    \caption{\textbf{Commensurate configurations with and without edge states. A, B}, The loss modulation of the periodic (commensurate) non-Hermitian topological models under investigation in our experiments for periodic models with and without edge states, respectively. It can be seen that the periodicity of rings' loss commensurate with the lattice periodicity for both cases. \textbf{C, D}, The schematic of the programmable integrated photonics platform for implementing the configurations with and without edge states, respectively. The programmable circuit is composed of a hexagonal mesh comprising reconfigurable interferometers and phase shifters. The inset details the arrangement of the multimode interferometers (MMIs) and phase shifters (PSs) forming the edges of the hexagons. The input port of each ring is indicated with the brown color. The designated output ports, shown with green/Magenta color, serve a dual purpose: they provide the prescribed loss for each ring and facilitate the measurement of stored energy within each ring. The equal coupling $t$ between adjacent rings is achieved by carefully adjusting the phase of the phase shifters within the common unit cell between rings. These unit cells, which control the coupling between rings, are highlighted in blue.}
    \label{fig:mesh}
\end{figure*}
We examine a one-dimensional non-Hermitian topological model, as theoretically
introduced in \cite{JosePPR2024}, wherein sites are interconnected with uniform coupling strengths, while exhibiting varying degrees of on site loss modulated by a sinusoidal function, leading to the following effective Hamiltonian:
\begin{align}
    H = \sum_{n=0}^{N-2}\left(t \hat{a}_n^{\dagger}\hat{a}_{n+1}+h.c. \right)
    -i\sum_{n=0}^{N-1}\left[v_1+v_2\sin\left(2\pi n \alpha+\delta \right) \right]\hat{a}_n^{\dagger}\hat{a}_n,
    \label{eq:Hamiltonian}
\end{align}
where $\hat{a}_n^{\dagger}$ and $\hat{a}_n$ are creation and annihilation operators in site $n$, $t$ is the coupling between neighboring sites, $\alpha$ is the inverse of the loss modulation period, $\delta$ represents the modulation phase, $v_2$ is the amplitude of the modulation, and $v_1$ is a uniform loss introduced to compensate the gain induced by sinusoidal gain/loss modulation.\\ 
A rational modulation frequency $\alpha = p/q$, where the number of sites $N$ is a multiple of $q$, ensures commensurability between the potential and the lattice periodicity, leading to a periodic scenario. Conversely, when the modulation frequency takes irrational values the potential is incommensurate with the lattice periodicity, resulting in aperiodic scenarios akin to quasicrystals.\\

To implement the non-Hermitian topological model experimentally, we utilize a programmable integrated photonics platform composed of a hexagonal mesh of reconfigurable Mach-Zehnder interferometers (MZI), as illustrated in Fig.\ref{fig:mesh}C (see supplementary note 1 for details). Recently, this platform has proven its potential to implement Hermitian topological Hamiltonians \cite{MehmetNatComm2024}. The inset shows the configuration of each MZI, composed of an input and an output 2x2 multimode interferometer (MMI) and a controllable phase shifter (PS) in each arm.  By adjusting the phase imparted by each PS we can configure the splitting ratio of each MZI, and thence the path followed by light in the mesh. Doing so, we configure the paths into an array of coupled ring resonators, using the MZIs common to two rings (cyan cells) to implement uniform coupling $t$. The loss in each ring is controlled by tapping the desired amount of power out of the mesh via one of the MZIs in each ring: green (magenta) cells highlight the MZIs used to tap a small (large) amount of power out of the ring, leading to a low (high) loss ring. These ports are connected to on-chip photodiodes that are used to monitor the power in each ring. Finally, we control the input excitation to each ring independently using the cells highlighted in brown. 
\begin{figure*}
    \centering
    \includegraphics[width=\textwidth]{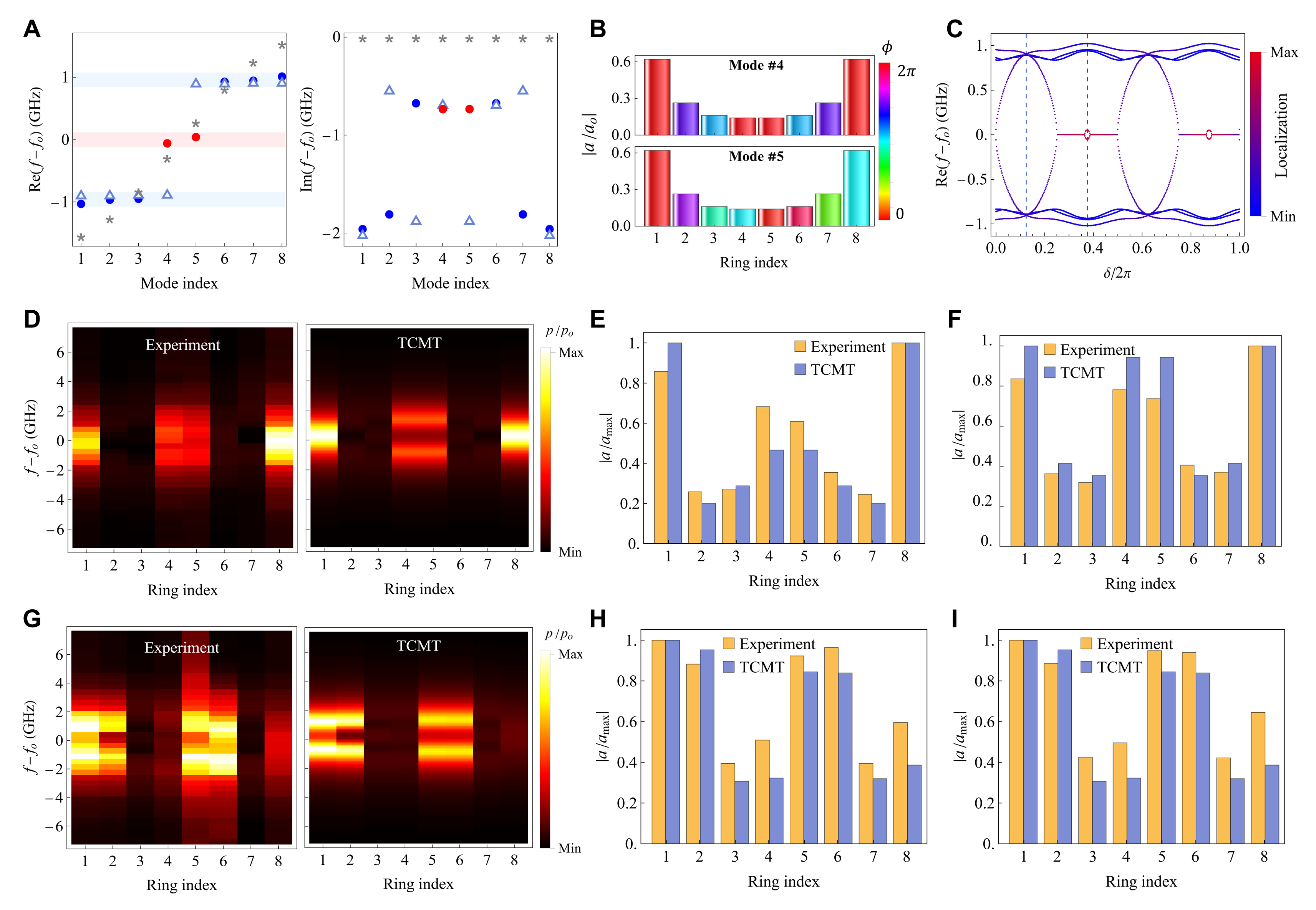}
    \caption{\textbf{Theoretical and experimental characterization of the commensurate configurations with/without edge states. A}, illustrates the real and imaginary parts of the eigenfrequencies for the commensurate case with edge states shown in Fig.\ref{fig:mesh}a (dot markers), commensurate case without edge states shown in Fig.\ref{fig:mesh}b (triangle markers), and the topologically trivial model devoid of any loss modulation (asterisk markers). \textbf{B}, Spatial distribution of edge states corresponding to the red points shown in Fig.\ref{fig:eigenvalues}A. The color bar represents the phase $\phi$ of the fields. \textbf{C}, The real part of the model's eigenfrequencies as a function of modulation phase, $\delta$, illustrating topological phase transitions. Points are color coded such that red indicates highly localized eigenmodes, while blue indicates highly delocalized ones. \textbf{D}, Local density of states (LDOS) spectrum obtained through experiment and temporal coupled mode theory (TCMT) analysis of the periodic system with edge states. 
    \textbf{E, F}, Normalized field amplitude at each ring for and excitation frequency of $f=f_o$ and $f=f_o+0.93$ GHz, respectively. \textbf{g}, LDOS spectrum for commensurate case without edge states (see Fig.\ref{fig:mesh}B). \textbf{H, I}, Normalized field amplitude at each ring for the excitation frequency of $f = f_o \pm 0.93$ GHz, respectively.}
    \label{fig:eigenvalues}
\end{figure*}
\subsection*{\textit{Configurations with commensurate loss modulation}}
We analyzed two distinct periodic (commensurate) scenarios with different topological features: one, illustrated in Fig.\ref{fig:mesh}A, exhibits edge states, while the other, depicted in Fig.\ref{fig:mesh}B, does not. Both periodic scenarios have a modulation period of $1/\alpha=4$, which corresponds to four sites (ring resonators) within each cell, and we consider a chain of two cells. The modulation phase parameter $\delta$ is set to $\delta=3\pi/4$ ($\delta=\pi/4$) for the periodic configuration with (without) edge states. In both cases, the periodicity of the sinusoidal loss function -- indicated by the dashed black line in Figs.\ref{fig:mesh}A and B -- aligns with the lattice periodicity.
The described commensurate scenarios were implemented  by reconfiguring the programmable mesh into a chain of eight rings with uniform coupling strength, $t=0.820$ GHz, and two discrete loss levels, $l_1=0.034$ GHz, $l_2=1.840$ GHz -- as shown in Figs.\ref{fig:mesh}C and D. These parameters relate to those in Eq. \eqref{eq:Hamiltonian} by the expressions $v_1=l_o+(l_1+l_2)/2$, and $v_2=(l_1-l_2)/\sqrt2$, where $l_o=0.35$GHz is the intrinsic decay rate of the ring resonators (see supplementary note 2 for details). Fig.\ref{fig:eigenvalues}A shows the real and imaginary parts of the eigenfrequencies of the systems implemented in Figs.\ref{fig:mesh}A (dot markers) and B (triangle markers). While both cases show the emergence of a band gap of approximately the same size, the first case shows two edge states, whose mode profiles are depicted in Fig.\ref{fig:eigenvalues}B, within the gap near the resonance frequency of the ring resonators at $f_o=193.366$ THz, highlighted in red, but the latter does not exhibit any edge modes. As it can be seen in the right panel of Fig.\ref{fig:mesh}A, each of those eigenvalues has a non-zero imaginary part (i.e. the eigenvalues are complex), a unique feature of non-Hermitian systems. For comparison, Fig.\ref{fig:eigenvalues}A also shows the eigenvalues of an equivalent system with the same uniform coupling but no loss modulation (asterisk markers). In stark contrast with the non-Hermitian case, this scenario does not exhibit any band gap or edge mode and possesses only real eigenvalues.
It is noteworthy that, the modulation phase parameter $\delta$ serves as a tuning mechanism for inducing topological phase transitions in the model \cite{BofengPRR2023}. Figure \ref{fig:eigenvalues}C illustrates the real part of the system's eigenfrequencies plotted against the modulation phase $\delta$. Each point on the plot is color-coded to represent the degree of localization of the corresponding eigenstate: red indicates high localization, while blue indicates high delocalization (see supplementary note 3 for details). We observe that, as the phase of modulation $\delta$ increases from zero, a band gap emerges, reaching its maximum around $\delta=\pi/4$. Subsequently, the band gap closes, signifying a topological phase transition, and the system eventually exhibits edge states, denoted by the red color points within the gap. Two distinct periodic configurations of the system, characterized by different modulation phase values under investigation in this study, are depicted by the dashed red ($\delta=3\pi/4$) and dashed blue ($\delta=\pi/4$) vertical lines in Fig.\ref{fig:eigenvalues}C.\\

First, we characterize the configuration featuring topological edge states by measuring the spectrum of its local density of states (LDOS), which provides valuable insights into both the eigenfrequencies and the spatial localization of the eigenmodes.
In Hermitian systems, the existence of the edge mode is probed by exciting the edge site with light at the frequency of the edge mode \cite{MehmetNatComm2024}. However, in non-Hermitian systems these conditions are not sufficient to excite the edge mode efficiently, and hence the LDOS is a more informative measurement 
(see supplementary note 4 for details). To measure the LDOS spectrum, we excited one ring at a time and measured its intracavity power as a function of excitation frequency. 
Fig.\ref{fig:eigenvalues}D displays the measured LDOS spectrum, normalized to the input power, exhibiting a notable level of agreement with the temporal coupled mode theory (TCMT) analysis. For excitation frequencies matching the eigenfrequency of the edge modes, i.e., modes \#4 and \#5 in Fig.\ref{fig:eigenvalues}A, the obtained LDOS spectrum exhibits pronounced intensity localization at the chain's edges, namely at ring\#1 and ring\#8. This localization pattern serves as a signature of the presence of edge states. To further illustrate the localization of the system's mode at $f=f_o$, Fig.\ref{fig:eigenvalues}E presents the normalized amplitude of the system's mode obtained from both experimental results and TCMT analysis by exciting each ring at a time and measuring its output power. Notably, this figure highlights the significant localization of the mode at the chain's edge rings.
The presence of the bulk modes in the band is illustrated by the relatively high intensity in the middle rings \#4 and \#5, peaking at frequencies $f-f_o\approx \pm 1$ GHz. To provide further clarity regarding the delocalization of these modes, we depicted the normalized field amplitude of the rings in Fig.\ref{fig:eigenvalues}F. Field amplitudes at each ring were acquired through excitation and subsequent local power detection for an excitation signal with a frequency difference of $f - f_o = 0.93$ GHz. Both the experimental results and those derived from TCMT analysis indicate that rings with similar levels of loss demonstrate comparable amplitudes for the mode, thus suggesting the mode's delocalization.
Next, we proceeded with similar LDOS analysis for the periodic configuration devoid of edge states, as illustrated in Fig.\ref{fig:mesh}B. The measured and TMCT calculated LDOS spectrum for this scenario is depicted in Fig.\ref{fig:eigenvalues}G. In this scenario, unlike the previous case featuring edge states, the LDOS spectrum does not exhibit a dominant intensity localized exclusively at the edge of the chain for any particular frequency. The heightened intensity around $f-f_o\approx \pm 1$ GHz is not confined solely to the rings at the chain's edge but is also spread across rings positioned in the middle of the chain which indicates a lack of the edge modes and the existence of delocalized modes in the bands. 
This finding is in agreement with the eigenfrequencies of the system plotted with triangles in Fig.\ref{fig:eigenvalues}A showing two bands around $f-f_o=\pm 0.9$ GHz. Furthermore, the normalized field amplitude at each ring, obtained through local excitation and power detection with an excitation frequencies of $f-f_o=\pm 0.93$ GHz, are shown in Figs.\ref{fig:eigenvalues}H and I, respectively, for both experimental and TCMT results. The comparable amplitudes observed for rings with similar loss values strongly corroborate our inference that the modes of the system in this configuration are delocalized.

\subsection*{\textit{Configuration with incommensurate loss modulation}}
\begin{figure*}
    \centering
    \includegraphics[width=\textwidth]{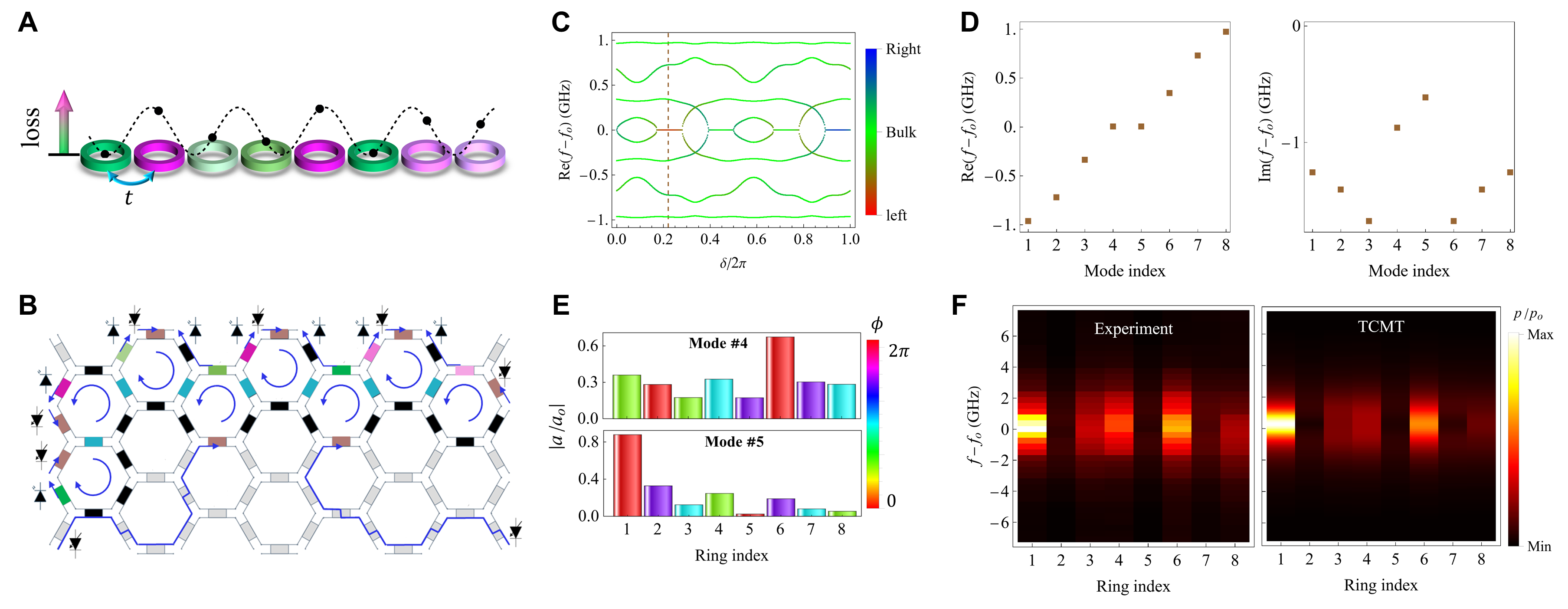}
    \caption{\textbf{A quasi-periodic (incommensurate) configuration. A}, The loss modulation for the incommensurate case is depicted by the dashed black line while the individual loss values for each ring are represented by black dots. As a result of the incommensurability of the loss modulation, the loss values for the rings do not repeat periodically. \textbf{B}, The schematic of the programmable integrated photonics platform designed for implementing the quasi-periodic configuration with incommensurate loss modulation for a 1D lattice of ring resonators. \textbf{C}, The real part of the eigenfrequencies of the quasi-periodic configuration with incommensurate loss modulation as a function of the modulation phase. Red (blue) points represent modes localized at the left (right) end of the lattice and green points represent delocalized modes. \textbf{D}, Real and imaginary part of the eigenfrequencies of the system with $\delta=0.44\pi$. \textbf{E}, Normalized mode profiles for eigenmodes \#4 and \#5 shown in \textbf{D}. The color bar shows the phase, $\phi$, of the fields. \textbf{F}, LDOS spectrum of the incommensurate configuration. The concentrated high power observed at ring \#1, along with the local maxima of power at ring \#6 for the excitation frequency $f = f_o$, serves as a signature of the coexistence of edge and bulk modes.}
    \label{fig:mesh_aperiodic}
\end{figure*}
Finally we turn our attention to the quasi-periodic case with incommensurate loss modulation, schematically depicted in Fig.\ref{fig:mesh_aperiodic}A. To achieve a quasi-periodic configuration we set the parameters as follows: $\alpha=(\sqrt{5}-1)/2$, $\delta=0.44\pi$, $t=0.63$ GHz, $v_1=l_o+0.94$ GHz, and $v_2=-0.92$ GHz in Eq. \ref{eq:Hamiltonian}. In Fig.\ref{fig:mesh_aperiodic}B, we show the implementation of this configuration on the programmable mesh. Notably, the parameter $\alpha$ being an irrational number leads to the values of ring losses (indicated by black points) not aligning with the lattice periodicity, resulting in an incommensurate (quasi-periodic) configuration. Once again, we plot the real part of the system's eigenfrequencies as a function of the modulation phase parameter in Fig.\ref{fig:mesh_aperiodic}C to elucidate the role of this parameter in topological phase transitions in the incommensurate case. This time we differentiate between localization at the left and right edges: red represents modes localized at the left end of the lattice, blue represents modes localized at the right end, and green represents delocalized modes (see supplementary note 3 for details). 
Here, the bands near the resonance frequency of the rings repeatedly intersect and open as the modulation phase varies. 
Fig.\ref{fig:mesh_aperiodic}D shows the real (left panel) and imaginary (right panel) parts of the system's eigenfrequencies for $\delta=0.44\pi$, corresponding to the dashed brown vertical line in Fig.\ref{fig:mesh_aperiodic}C. Unlike the commensurate scenario, where two edge states are present at $f\approx f_o$, here only one mode exhibits localization at the lattice's edge. This becomes evident in Fig.\ref{fig:mesh_aperiodic}E, depicting the field amplitude inside the rings for these two modes (\#4 and \#5). The origin of this is in the fact that, despite the real part of their eigenfrequencies being identical, their imaginary parts differ, indicating non-degeneracy between them. In non-Hermitian systems like the one in our experiment, two modes with the same real part but different imaginary part of their eigenfrequencies can exhibit vastly different localization behavior.
The presence of a single edge mode is easily rationalized from the pumping of the edge mode as a function of the phason. This feature, analogous to the pumping of the edge modes in the Hermitian AAH model, establishes that different terminations of the system are associated to sampling different values of the phason. A single experimental configuration, thus corresponds to a phason $\delta$ featuring a mode on a single edge (see supplementary note 5 for further details). 
Finally, it is worth noting that this incommensurate scenario will provide an attractive platform to observe localization-delocalization transitions in non-Hermitian systems for longer chains.\\

The LDOS spectrum for the quasi-periodic chain is depicted in Fig.\ref{fig:mesh_aperiodic}F, again, in good agreement with TCMT. Notably, in the obtained LDOS spectrum, ring \#1 demonstrates significant localization at $f=f_o$, while intermediate rings also exhibit considerable power at the same frequency. This observation suggests the coexistence of a localized and a delocalized state at $f=f_o$. 
\subsection*{\textit{Robustness study of the edge states}}
\begin{figure*}
    \centering
    \includegraphics[width=\textwidth]{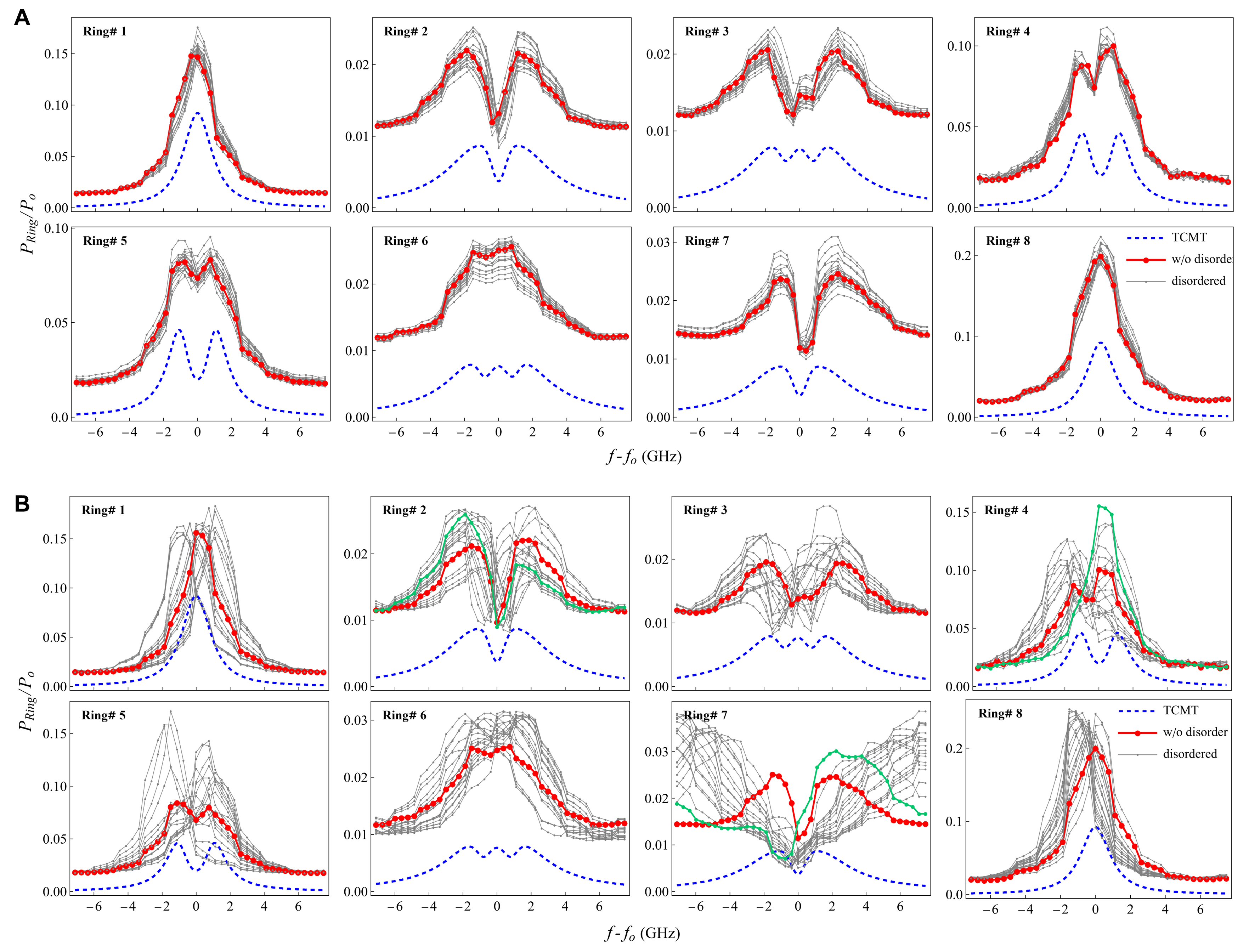}
    \caption{\textbf{Robustness against disorder in the loss and resonance frequency in the commensurate case. A, B} Plots of local power spectrum for each ring resonator of the commensurate configuration featuring edge states depicted in Fig.\ref{fig:mesh} \textbf{A}. The local power spectrum of each ring was obtained by exciting it and subsequently measuring its power while varying the frequency of the exciting signal. The red curves represent the local power spectrum without any introduced perturbation to the system, acting as a reference for comparison with the dashed blue curves obtained through TCMT analysis of the system. The gray curves represent the local power spectrum after introducing a Gaussian disorder to the loss of each ring (panel \textbf{A}) and a uniform disorder to the resonance frequency of each ring (panel \textbf{B} ). Notably, the disorder in the loss neither modifies the shape of the spectrum nor disrupts their frequency dependence. Instead, it introduces fluctuations in the power of the rings. However, disorders in rings' resonance frequency not only perturbs the frequency dependence of the spectrum but also modifies its shape, as demonstrated in selected plots by the inclusion of green curves for enhanced clarity.}
    \label{fig:random_loss_edge}
\end{figure*}

We first investigate the robustness of the edge states in the periodic configuration depicted in Fig.\ref{fig:mesh}A. The resilience of the system against two kind of disorders, namely variations in the loss and variations in the resonance frequency of the ring resonators, has been examined. For each excitation frequency, we deliberately perturbed the decay rates, $l_{1,2}$, of the rings, with perturbations following a Gaussian distribution having a standard deviation of $\sigma_{1,2}=0.1\times l_{1,2}$. Each excitation frequency underwent measurements with 20 different random iterations of disorder. In Fig.\ref{fig:random_loss_edge}A, we show the normalized measured power in each ring as a function of frequency under local excitation: the red curves show the measurements in the absence of disorder and the grey curves show the measurements in the presence of each random iteration of disorder in ring's decay rates. For reference, the power spectrum obtained by TCMT is plotted with dashed blue curves. Disorders in the loss respect the particle-hole symmetry of the system \cite{JosePPR2024,BofengPRR2023}, thereby resulting in the robustness of the topological states against loss disorder.

In contrast, disorder in the resonance frequency of the rings, or in the rings couplings, does not respect the particle-hole symmetry leading to the disappearance of the robustness of the edge states. To demonstrate this we introduce deliberate perturbations in the resonance frequency of the ring resonators with uniform distribution and standard deviation of $\sigma=0.04\times \text{FSR}$, where $\text{FSR}=13.97$ GHz is the free spectral range of the ring resonators and performed measurements of the power spectrum per ring as described above. The experimental results are shown in Fig.\ref{fig:random_loss_edge}B: in the absence of disorder in the resonance frequencies (red curve), the experimental results align closely with the TCMT results (dashed blue curve); in contrast, the gray curves, representing the outcomes under 20 random iterations of disorder in the resonance frequencies show significant fluctuations in the power spectra, observed not only in the vertical direction but also horizontally. This behavior can be contrasted with the effects of loss disorders shown in Fig.\ref{fig:random_loss_edge}A, which primarily induce vertical fluctuations in the spectra. Another significant observation evident in Fig.\ref{fig:random_loss_edge}B is the alteration in the shape of the rings' power spectrum upon introducing perturbations at the resonance frequency of the rings. To emphasize this further, we have plotted some power spectra with green curves to magnify the change in the shape of the power spectrum. These observations highlight the susceptibility of edge states to disorder in the resonance frequency of the rings, or equivalently, to variations in the coupling between rings.

\begin{figure*}
    \centering
    \includegraphics[width=\textwidth]{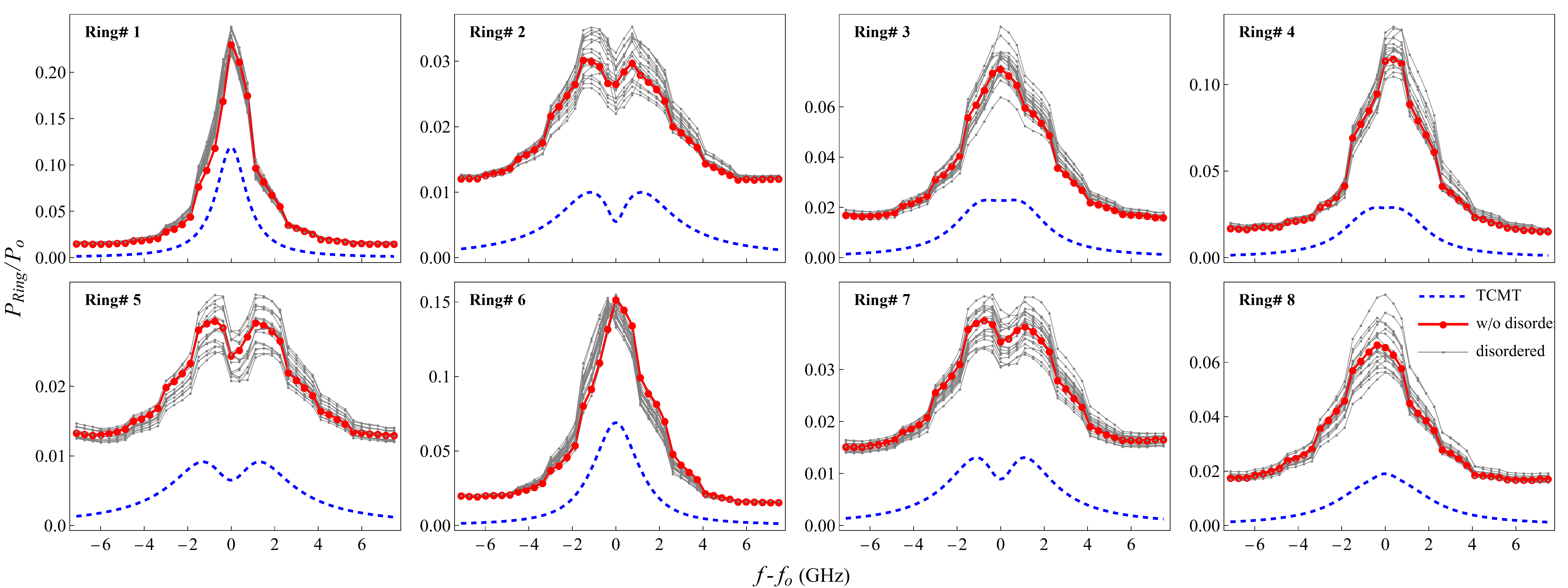}
    \caption{\textbf{Robustness against disorder in the loss in the incommensurate case.} Plots of local power spectrum for each ring resonator of the incommensurate configuration depicted in Fig.\ref{fig:mesh_aperiodic}A. The red curves represent the local power spectrum without any introduced perturbation to the system, which should be compared with the dashed blue curves obtained through TCMT analysis of the system. The gray curves represent the local power spectrum after introducing a Gaussian disorder to the loss of each ring.}
    \label{fig:random_loss_incommensurate}
\end{figure*}

We performed a similar analysis of the robustness of the edge mode in the incommensurate scenario. Figure \ref{fig:random_loss_incommensurate} shows  measured power spectra of each ring in the absence of disorder (red curves) and under 20 different iterations of random disorder in the loss drawn from a Gaussian distribution with $\sigma_n = 0.1\times l_n$, $n=1,2,\cdots,8$. These results highlight a high level of robustness to disorder preserving the particle-holes symmetry, similar to that of the periodic case with edge states. 

\section*{Discussion and Conclusion}
In our work, we have demonstrated a photonic system where the topological properties arise, exclusively, from the system's non-Hermiticity. We showed that nontrivial topological phases and topological edge modes can arise in the presence of periodic or quasi-periodic loss modulation. Our robustness analysis revealed that the topological edge modes in these systems are immune to kinds of disorder that preserve the particle-hole symmetry, such as disorder in the loss, but not to disorder that breaks such symmetry, such as disorder in the resonant frequency that affects the coupling terms.\\
The AAH model \cite{AubryAndre1980} has been extensively studied in the realm of condensed matter physics, providing insights into the behavior of quantum particles in periodic or quasi-periodic potentials. In photonics, the AAH model has underpinned many of the studies of topology in quasicrystals, leading to crucial demonstrations of topological pumping \cite{ZilberbergPRB2015} and topologically protected quantum interference \cite{TambascoScience2018} among many others. Adding to the previous literature on non-Hermitian extensions of the AAH model \cite{YucePLA2014,LonghiPRB2019,ZengPRB2020,weidemann2022topological,TakataPRL2018,BofengPRR2023,JosePPR2024}, our programmable non-Hermitian platform introduces a new angle of observation, allowing for extensive exploration of the intricate dynamics in non-Hermitian systems.
The quasiperiodic configuration of this model, specifically, offers a convenient testing ground to explore phase transitions in dissipative quasicrystals.\\
Overall our results point to new avenues of research in non-Hermitian topological photonics where loss, far from being a nuisance, becomes a powerful knob to control the topological properties of electromagnetic modes.

\section*{Data availability}
The data that support the findings of this study are available from the corresponding authors upon reasonable request.

\bibliography{Reference}

\begin{thebibliography}{41}%
\makeatletter
\providecommand \@ifxundefined [1]{%
 \@ifx{#1\undefined}
}%
\providecommand \@ifnum [1]{%
 \ifnum #1\expandafter \@firstoftwo
 \else \expandafter \@secondoftwo
 \fi
}%
\providecommand \@ifx [1]{%
 \ifx #1\expandafter \@firstoftwo
 \else \expandafter \@secondoftwo
 \fi
}%
\providecommand \natexlab [1]{#1}%
\providecommand \enquote  [1]{``#1''}%
\providecommand \bibnamefont  [1]{#1}%
\providecommand \bibfnamefont [1]{#1}%
\providecommand \citenamefont [1]{#1}%
\providecommand \href@noop [0]{\@secondoftwo}%
\providecommand \href [0]{\begingroup \@sanitize@url \@href}%
\providecommand \@href[1]{\@@startlink{#1}\@@href}%
\providecommand \@@href[1]{\endgroup#1\@@endlink}%
\providecommand \@sanitize@url [0]{\catcode `\\12\catcode `\$12\catcode
  `\&12\catcode `\#12\catcode `\^12\catcode `\_12\catcode `\%12\relax}%
\providecommand \@@startlink[1]{}%
\providecommand \@@endlink[0]{}%
\providecommand \url  [0]{\begingroup\@sanitize@url \@url }%
\providecommand \@url [1]{\endgroup\@href {#1}{\urlprefix }}%
\providecommand \urlprefix  [0]{URL }%
\providecommand \Eprint [0]{\href }%
\providecommand \doibase [0]{https://doi.org/}%
\providecommand \selectlanguage [0]{\@gobble}%
\providecommand \bibinfo  [0]{\@secondoftwo}%
\providecommand \bibfield  [0]{\@secondoftwo}%
\providecommand \translation [1]{[#1]}%
\providecommand \BibitemOpen [0]{}%
\providecommand \bibitemStop [0]{}%
\providecommand \bibitemNoStop [0]{.\EOS\space}%
\providecommand \EOS [0]{\spacefactor3000\relax}%
\providecommand \BibitemShut  [1]{\csname bibitem#1\endcsname}%
\let\auto@bib@innerbib\@empty
\bibitem [{\citenamefont {Ozawa}\ \emph {et~al.}(2019)\citenamefont {Ozawa},
  \citenamefont {Price}, \citenamefont {Amo}, \citenamefont {Goldman},
  \citenamefont {Hafezi}, \citenamefont {Lu}, \citenamefont {Rechtsman},
  \citenamefont {Schuster}, \citenamefont {Simon}, \citenamefont {Zilberberg}
  \emph {et~al.}}]{ozawa2019topological}%
  \BibitemOpen
  \bibfield  {author} {\bibinfo {author} {\bibfnamefont {T.}~\bibnamefont
  {Ozawa}}, \bibinfo {author} {\bibfnamefont {H.~M.}\ \bibnamefont {Price}},
  \bibinfo {author} {\bibfnamefont {A.}~\bibnamefont {Amo}}, \bibinfo {author}
  {\bibfnamefont {N.}~\bibnamefont {Goldman}}, \bibinfo {author} {\bibfnamefont
  {M.}~\bibnamefont {Hafezi}}, \bibinfo {author} {\bibfnamefont
  {L.}~\bibnamefont {Lu}}, \bibinfo {author} {\bibfnamefont {M.~C.}\
  \bibnamefont {Rechtsman}}, \bibinfo {author} {\bibfnamefont {D.}~\bibnamefont
  {Schuster}}, \bibinfo {author} {\bibfnamefont {J.}~\bibnamefont {Simon}},
  \bibinfo {author} {\bibfnamefont {O.}~\bibnamefont {Zilberberg}}, \emph
  {et~al.},\ }\bibfield  {title} {\bibinfo {title} {Topological photonics},\
  }\href@noop {} {\bibfield  {journal} {\bibinfo  {journal} {Reviews of Modern
  Physics}\ }\textbf {\bibinfo {volume} {91}},\ \bibinfo {pages} {015006}
  (\bibinfo {year} {2019})}\BibitemShut {NoStop}%
\bibitem [{\citenamefont {Price}\ \emph {et~al.}(2022)\citenamefont {Price},
  \citenamefont {Chong}, \citenamefont {Khanikaev}, \citenamefont {Schomerus},
  \citenamefont {Maczewsky}, \citenamefont {Kremer}, \citenamefont {Heinrich},
  \citenamefont {Szameit}, \citenamefont {Zilberberg}, \citenamefont {Yang}
  \emph {et~al.}}]{price2022roadmap}%
  \BibitemOpen
  \bibfield  {author} {\bibinfo {author} {\bibfnamefont {H.}~\bibnamefont
  {Price}}, \bibinfo {author} {\bibfnamefont {Y.}~\bibnamefont {Chong}},
  \bibinfo {author} {\bibfnamefont {A.}~\bibnamefont {Khanikaev}}, \bibinfo
  {author} {\bibfnamefont {H.}~\bibnamefont {Schomerus}}, \bibinfo {author}
  {\bibfnamefont {L.~J.}\ \bibnamefont {Maczewsky}}, \bibinfo {author}
  {\bibfnamefont {M.}~\bibnamefont {Kremer}}, \bibinfo {author} {\bibfnamefont
  {M.}~\bibnamefont {Heinrich}}, \bibinfo {author} {\bibfnamefont
  {A.}~\bibnamefont {Szameit}}, \bibinfo {author} {\bibfnamefont
  {O.}~\bibnamefont {Zilberberg}}, \bibinfo {author} {\bibfnamefont
  {Y.}~\bibnamefont {Yang}}, \emph {et~al.},\ }\bibfield  {title} {\bibinfo
  {title} {Roadmap on topological photonics},\ }\href@noop {} {\bibfield
  {journal} {\bibinfo  {journal} {Journal of Physics: Photonics}\ }\textbf
  {\bibinfo {volume} {4}},\ \bibinfo {pages} {032501} (\bibinfo {year}
  {2022})}\BibitemShut {NoStop}%
\bibitem [{\citenamefont {Bandres}\ \emph {et~al.}(2018)\citenamefont
  {Bandres}, \citenamefont {Wittek}, \citenamefont {Harari}, \citenamefont
  {Parto}, \citenamefont {Ren}, \citenamefont {Segev}, \citenamefont
  {Christodoulides},\ and\ \citenamefont
  {Khajavikhan}}]{bandres2018topological}%
  \BibitemOpen
  \bibfield  {author} {\bibinfo {author} {\bibfnamefont {M.~A.}\ \bibnamefont
  {Bandres}}, \bibinfo {author} {\bibfnamefont {S.}~\bibnamefont {Wittek}},
  \bibinfo {author} {\bibfnamefont {G.}~\bibnamefont {Harari}}, \bibinfo
  {author} {\bibfnamefont {M.}~\bibnamefont {Parto}}, \bibinfo {author}
  {\bibfnamefont {J.}~\bibnamefont {Ren}}, \bibinfo {author} {\bibfnamefont
  {M.}~\bibnamefont {Segev}}, \bibinfo {author} {\bibfnamefont {D.~N.}\
  \bibnamefont {Christodoulides}},\ and\ \bibinfo {author} {\bibfnamefont
  {M.}~\bibnamefont {Khajavikhan}},\ }\bibfield  {title} {\bibinfo {title}
  {Topological insulator laser: Experiments},\ }\href@noop {} {\bibfield
  {journal} {\bibinfo  {journal} {Science}\ }\textbf {\bibinfo {volume}
  {359}},\ \bibinfo {pages} {eaar4005} (\bibinfo {year} {2018})}\BibitemShut
  {NoStop}%
\bibitem [{\citenamefont {Contractor}\ \emph {et~al.}(2022)\citenamefont
  {Contractor}, \citenamefont {Noh}, \citenamefont {Redjem}, \citenamefont
  {Qarony}, \citenamefont {Martin}, \citenamefont {Dhuey}, \citenamefont
  {Schwartzberg},\ and\ \citenamefont {Kant{\'e}}}]{contractor2022scalable}%
  \BibitemOpen
  \bibfield  {author} {\bibinfo {author} {\bibfnamefont {R.}~\bibnamefont
  {Contractor}}, \bibinfo {author} {\bibfnamefont {W.}~\bibnamefont {Noh}},
  \bibinfo {author} {\bibfnamefont {W.}~\bibnamefont {Redjem}}, \bibinfo
  {author} {\bibfnamefont {W.}~\bibnamefont {Qarony}}, \bibinfo {author}
  {\bibfnamefont {E.}~\bibnamefont {Martin}}, \bibinfo {author} {\bibfnamefont
  {S.}~\bibnamefont {Dhuey}}, \bibinfo {author} {\bibfnamefont
  {A.}~\bibnamefont {Schwartzberg}},\ and\ \bibinfo {author} {\bibfnamefont
  {B.}~\bibnamefont {Kant{\'e}}},\ }\bibfield  {title} {\bibinfo {title}
  {Scalable single-mode surface-emitting laser via open-dirac singularities},\
  }\href@noop {} {\bibfield  {journal} {\bibinfo  {journal} {Nature}\ }\textbf
  {\bibinfo {volume} {608}},\ \bibinfo {pages} {692} (\bibinfo {year}
  {2022})}\BibitemShut {NoStop}%
\bibitem [{\citenamefont {Kumar}\ \emph {et~al.}(2022)\citenamefont {Kumar},
  \citenamefont {Gupta}, \citenamefont {Pitchappa}, \citenamefont {Tan},
  \citenamefont {Wang},\ and\ \citenamefont {Singh}}]{kumar2022topological}%
  \BibitemOpen
  \bibfield  {author} {\bibinfo {author} {\bibfnamefont {A.}~\bibnamefont
  {Kumar}}, \bibinfo {author} {\bibfnamefont {M.}~\bibnamefont {Gupta}},
  \bibinfo {author} {\bibfnamefont {P.}~\bibnamefont {Pitchappa}}, \bibinfo
  {author} {\bibfnamefont {Y.~J.}\ \bibnamefont {Tan}}, \bibinfo {author}
  {\bibfnamefont {N.}~\bibnamefont {Wang}},\ and\ \bibinfo {author}
  {\bibfnamefont {R.}~\bibnamefont {Singh}},\ }\bibfield  {title} {\bibinfo
  {title} {Topological sensor on a silicon chip},\ }\href@noop {} {\bibfield
  {journal} {\bibinfo  {journal} {Applied Physics Letters}\ }\textbf {\bibinfo
  {volume} {121}} (\bibinfo {year} {2022})}\BibitemShut {NoStop}%
\bibitem [{\citenamefont {Blanco-Redondo}\ \emph {et~al.}(2018)\citenamefont
  {Blanco-Redondo}, \citenamefont {Bell}, \citenamefont {Oren}, \citenamefont
  {Eggleton},\ and\ \citenamefont {Segev}}]{blanco2018topological}%
  \BibitemOpen
  \bibfield  {author} {\bibinfo {author} {\bibfnamefont {A.}~\bibnamefont
  {Blanco-Redondo}}, \bibinfo {author} {\bibfnamefont {B.}~\bibnamefont
  {Bell}}, \bibinfo {author} {\bibfnamefont {D.}~\bibnamefont {Oren}}, \bibinfo
  {author} {\bibfnamefont {B.~J.}\ \bibnamefont {Eggleton}},\ and\ \bibinfo
  {author} {\bibfnamefont {M.}~\bibnamefont {Segev}},\ }\bibfield  {title}
  {\bibinfo {title} {Topological protection of biphoton states},\ }\href@noop
  {} {\bibfield  {journal} {\bibinfo  {journal} {Science}\ }\textbf {\bibinfo
  {volume} {362}},\ \bibinfo {pages} {568} (\bibinfo {year}
  {2018})}\BibitemShut {NoStop}%
\bibitem [{\citenamefont {Mittal}\ \emph {et~al.}(2018)\citenamefont {Mittal},
  \citenamefont {Goldschmidt},\ and\ \citenamefont
  {Hafezi}}]{mittal2018topological}%
  \BibitemOpen
  \bibfield  {author} {\bibinfo {author} {\bibfnamefont {S.}~\bibnamefont
  {Mittal}}, \bibinfo {author} {\bibfnamefont {E.~A.}\ \bibnamefont
  {Goldschmidt}},\ and\ \bibinfo {author} {\bibfnamefont {M.}~\bibnamefont
  {Hafezi}},\ }\bibfield  {title} {\bibinfo {title} {A topological source of
  quantum light},\ }\href@noop {} {\bibfield  {journal} {\bibinfo  {journal}
  {Nature}\ }\textbf {\bibinfo {volume} {561}},\ \bibinfo {pages} {502}
  (\bibinfo {year} {2018})}\BibitemShut {NoStop}%
\bibitem [{\citenamefont {Dai}\ \emph {et~al.}(2022)\citenamefont {Dai},
  \citenamefont {Ao}, \citenamefont {Bao}, \citenamefont {Mao}, \citenamefont
  {Chi}, \citenamefont {Fu}, \citenamefont {You}, \citenamefont {Chen},
  \citenamefont {Zhai}, \citenamefont {Tang} \emph
  {et~al.}}]{dai2022topologically}%
  \BibitemOpen
  \bibfield  {author} {\bibinfo {author} {\bibfnamefont {T.}~\bibnamefont
  {Dai}}, \bibinfo {author} {\bibfnamefont {Y.}~\bibnamefont {Ao}}, \bibinfo
  {author} {\bibfnamefont {J.}~\bibnamefont {Bao}}, \bibinfo {author}
  {\bibfnamefont {J.}~\bibnamefont {Mao}}, \bibinfo {author} {\bibfnamefont
  {Y.}~\bibnamefont {Chi}}, \bibinfo {author} {\bibfnamefont {Z.}~\bibnamefont
  {Fu}}, \bibinfo {author} {\bibfnamefont {Y.}~\bibnamefont {You}}, \bibinfo
  {author} {\bibfnamefont {X.}~\bibnamefont {Chen}}, \bibinfo {author}
  {\bibfnamefont {C.}~\bibnamefont {Zhai}}, \bibinfo {author} {\bibfnamefont
  {B.}~\bibnamefont {Tang}}, \emph {et~al.},\ }\bibfield  {title} {\bibinfo
  {title} {Topologically protected quantum entanglement emitters},\ }\href@noop
  {} {\bibfield  {journal} {\bibinfo  {journal} {Nature Photonics}\ }\textbf
  {\bibinfo {volume} {16}},\ \bibinfo {pages} {248} (\bibinfo {year}
  {2022})}\BibitemShut {NoStop}%
\bibitem [{\citenamefont {Nasari}\ \emph {et~al.}(2023)\citenamefont {Nasari},
  \citenamefont {Pyrialakos}, \citenamefont {Christodoulides},\ and\
  \citenamefont {Khajavikhan}}]{Nasari:23}%
  \BibitemOpen
  \bibfield  {author} {\bibinfo {author} {\bibfnamefont {H.}~\bibnamefont
  {Nasari}}, \bibinfo {author} {\bibfnamefont {G.~G.}\ \bibnamefont
  {Pyrialakos}}, \bibinfo {author} {\bibfnamefont {D.~N.}\ \bibnamefont
  {Christodoulides}},\ and\ \bibinfo {author} {\bibfnamefont {M.}~\bibnamefont
  {Khajavikhan}},\ }\bibfield  {title} {\bibinfo {title} {Non-hermitian
  topological photonics},\ }\href {https://doi.org/10.1364/OME.483361}
  {\bibfield  {journal} {\bibinfo  {journal} {Opt. Mater. Express}\ }\textbf
  {\bibinfo {volume} {13}},\ \bibinfo {pages} {870} (\bibinfo {year}
  {2023})}\BibitemShut {NoStop}%
\bibitem [{\citenamefont {Yan}\ \emph {et~al.}(2023)\citenamefont {Yan},
  \citenamefont {Zhao}, \citenamefont {Zhou}, \citenamefont {Ma}, \citenamefont
  {Lyu}, \citenamefont {Chu}, \citenamefont {Hu},\ and\ \citenamefont
  {Gong}}]{Yan_NanPhot2023}%
  \BibitemOpen
  \bibfield  {author} {\bibinfo {author} {\bibfnamefont {Q.}~\bibnamefont
  {Yan}}, \bibinfo {author} {\bibfnamefont {B.}~\bibnamefont {Zhao}}, \bibinfo
  {author} {\bibfnamefont {R.}~\bibnamefont {Zhou}}, \bibinfo {author}
  {\bibfnamefont {R.}~\bibnamefont {Ma}}, \bibinfo {author} {\bibfnamefont
  {Q.}~\bibnamefont {Lyu}}, \bibinfo {author} {\bibfnamefont {S.}~\bibnamefont
  {Chu}}, \bibinfo {author} {\bibfnamefont {X.}~\bibnamefont {Hu}},\ and\
  \bibinfo {author} {\bibfnamefont {Q.}~\bibnamefont {Gong}},\ }\bibfield
  {title} {\bibinfo {title} {Advances and applications on non-hermitian
  topological photonics},\ }\href
  {https://doi.org/doi:10.1515/nanoph-2022-0775} {\bibfield  {journal}
  {\bibinfo  {journal} {Nanophotonics}\ }\textbf {\bibinfo {volume} {12}},\
  \bibinfo {pages} {2247} (\bibinfo {year} {2023})}\BibitemShut {NoStop}%
\bibitem [{\citenamefont {Kawabata}\ \emph {et~al.}(2019)\citenamefont
  {Kawabata}, \citenamefont {Shiozaki}, \citenamefont {Ueda},\ and\
  \citenamefont {Sato}}]{Kawabata_PhysRevX2019}%
  \BibitemOpen
  \bibfield  {author} {\bibinfo {author} {\bibfnamefont {K.}~\bibnamefont
  {Kawabata}}, \bibinfo {author} {\bibfnamefont {K.}~\bibnamefont {Shiozaki}},
  \bibinfo {author} {\bibfnamefont {M.}~\bibnamefont {Ueda}},\ and\ \bibinfo
  {author} {\bibfnamefont {M.}~\bibnamefont {Sato}},\ }\bibfield  {title}
  {\bibinfo {title} {Symmetry and topology in non-hermitian physics},\ }\href
  {https://doi.org/10.1103/PhysRevX.9.041015} {\bibfield  {journal} {\bibinfo
  {journal} {Phys. Rev. X}\ }\textbf {\bibinfo {volume} {9}},\ \bibinfo {pages}
  {041015} (\bibinfo {year} {2019})}\BibitemShut {NoStop}%
\bibitem [{\citenamefont {Meng}\ \emph {et~al.}(2024)\citenamefont {Meng},
  \citenamefont {Ang},\ and\ \citenamefont {Lee}}]{meng2024exceptional}%
  \BibitemOpen
  \bibfield  {author} {\bibinfo {author} {\bibfnamefont {H.}~\bibnamefont
  {Meng}}, \bibinfo {author} {\bibfnamefont {Y.~S.}\ \bibnamefont {Ang}},\ and\
  \bibinfo {author} {\bibfnamefont {C.~H.}\ \bibnamefont {Lee}},\ }\bibfield
  {title} {\bibinfo {title} {Exceptional points in non-hermitian systems:
  Applications and recent developments},\ }\href@noop {} {\bibfield  {journal}
  {\bibinfo  {journal} {Applied Physics Letters}\ }\textbf {\bibinfo {volume}
  {124}} (\bibinfo {year} {2024})}\BibitemShut {NoStop}%
\bibitem [{\citenamefont {Ding}\ \emph {et~al.}(2022)\citenamefont {Ding},
  \citenamefont {Fang},\ and\ \citenamefont {Ma}}]{ding2022non}%
  \BibitemOpen
  \bibfield  {author} {\bibinfo {author} {\bibfnamefont {K.}~\bibnamefont
  {Ding}}, \bibinfo {author} {\bibfnamefont {C.}~\bibnamefont {Fang}},\ and\
  \bibinfo {author} {\bibfnamefont {G.}~\bibnamefont {Ma}},\ }\bibfield
  {title} {\bibinfo {title} {Non-hermitian topology and exceptional-point
  geometries},\ }\href@noop {} {\bibfield  {journal} {\bibinfo  {journal}
  {Nature Reviews Physics}\ }\textbf {\bibinfo {volume} {4}},\ \bibinfo {pages}
  {745} (\bibinfo {year} {2022})}\BibitemShut {NoStop}%
\bibitem [{\citenamefont {Rudner}\ and\ \citenamefont
  {Levitov}(2009)}]{LevitovPRL2009}%
  \BibitemOpen
  \bibfield  {author} {\bibinfo {author} {\bibfnamefont {M.~S.}\ \bibnamefont
  {Rudner}}\ and\ \bibinfo {author} {\bibfnamefont {L.~S.}\ \bibnamefont
  {Levitov}},\ }\bibfield  {title} {\bibinfo {title} {Topological transition in
  a non-hermitian quantum walk},\ }\href
  {https://doi.org/10.1103/PhysRevLett.102.065703} {\bibfield  {journal}
  {\bibinfo  {journal} {Phys. Rev. Lett.}\ }\textbf {\bibinfo {volume} {102}},\
  \bibinfo {pages} {065703} (\bibinfo {year} {2009})}\BibitemShut {NoStop}%
\bibitem [{\citenamefont {Esaki}\ \emph {et~al.}(2011)\citenamefont {Esaki},
  \citenamefont {Sato}, \citenamefont {Hasebe},\ and\ \citenamefont
  {Kohmoto}}]{KohmotoPRB2011}%
  \BibitemOpen
  \bibfield  {author} {\bibinfo {author} {\bibfnamefont {K.}~\bibnamefont
  {Esaki}}, \bibinfo {author} {\bibfnamefont {M.}~\bibnamefont {Sato}},
  \bibinfo {author} {\bibfnamefont {K.}~\bibnamefont {Hasebe}},\ and\ \bibinfo
  {author} {\bibfnamefont {M.}~\bibnamefont {Kohmoto}},\ }\bibfield  {title}
  {\bibinfo {title} {Edge states and topological phases in non-hermitian
  systems},\ }\href {https://doi.org/10.1103/PhysRevB.84.205128} {\bibfield
  {journal} {\bibinfo  {journal} {Phys. Rev. B}\ }\textbf {\bibinfo {volume}
  {84}},\ \bibinfo {pages} {205128} (\bibinfo {year} {2011})}\BibitemShut
  {NoStop}%
\bibitem [{\citenamefont {Diehl}\ \emph {et~al.}(2011)\citenamefont {Diehl},
  \citenamefont {Rico}, \citenamefont {Baranov},\ and\ \citenamefont
  {Zoller}}]{ZollerNatPhys2011}%
  \BibitemOpen
  \bibfield  {author} {\bibinfo {author} {\bibfnamefont {S.}~\bibnamefont
  {Diehl}}, \bibinfo {author} {\bibfnamefont {E.}~\bibnamefont {Rico}},
  \bibinfo {author} {\bibfnamefont {M.~A.}\ \bibnamefont {Baranov}},\ and\
  \bibinfo {author} {\bibfnamefont {P.}~\bibnamefont {Zoller}},\ }\bibfield
  {title} {\bibinfo {title} {Topology by dissipation in atomic quantum wires},\
  }\href {https://doi.org/10.1038/nphys2106} {\bibfield  {journal} {\bibinfo
  {journal} {Nat. Phys.}\ }\textbf {\bibinfo {volume} {7}},\ \bibinfo {pages}
  {971} (\bibinfo {year} {2011})}\BibitemShut {NoStop}%
\bibitem [{\citenamefont {Schomerus}(2013)}]{schomerus2013topologically}%
  \BibitemOpen
  \bibfield  {author} {\bibinfo {author} {\bibfnamefont {H.}~\bibnamefont
  {Schomerus}},\ }\bibfield  {title} {\bibinfo {title} {Topologically protected
  midgap states in complex photonic lattices},\ }\href@noop {} {\bibfield
  {journal} {\bibinfo  {journal} {Optics letters}\ }\textbf {\bibinfo {volume}
  {38}},\ \bibinfo {pages} {1912} (\bibinfo {year} {2013})}\BibitemShut
  {NoStop}%
\bibitem [{\citenamefont {Leykam}\ \emph {et~al.}(2017)\citenamefont {Leykam},
  \citenamefont {Bliokh}, \citenamefont {Huang}, \citenamefont {Chong},\ and\
  \citenamefont {Nori}}]{leykam2017edge}%
  \BibitemOpen
  \bibfield  {author} {\bibinfo {author} {\bibfnamefont {D.}~\bibnamefont
  {Leykam}}, \bibinfo {author} {\bibfnamefont {K.~Y.}\ \bibnamefont {Bliokh}},
  \bibinfo {author} {\bibfnamefont {C.}~\bibnamefont {Huang}}, \bibinfo
  {author} {\bibfnamefont {Y.~D.}\ \bibnamefont {Chong}},\ and\ \bibinfo
  {author} {\bibfnamefont {F.}~\bibnamefont {Nori}},\ }\bibfield  {title}
  {\bibinfo {title} {Edge modes, degeneracies, and topological numbers in
  non-hermitian systems},\ }\href@noop {} {\bibfield  {journal} {\bibinfo
  {journal} {Physical review letters}\ }\textbf {\bibinfo {volume} {118}},\
  \bibinfo {pages} {040401} (\bibinfo {year} {2017})}\BibitemShut {NoStop}%
\bibitem [{\citenamefont {Res{\'e}ndiz-V{\'a}zquez}\ \emph
  {et~al.}(2020)\citenamefont {Res{\'e}ndiz-V{\'a}zquez}, \citenamefont
  {Tschernig}, \citenamefont {Perez-Leija}, \citenamefont {Busch},\ and\
  \citenamefont {Le{\'o}n-Montiel}}]{resendiz2020topological}%
  \BibitemOpen
  \bibfield  {author} {\bibinfo {author} {\bibfnamefont {P.}~\bibnamefont
  {Res{\'e}ndiz-V{\'a}zquez}}, \bibinfo {author} {\bibfnamefont
  {K.}~\bibnamefont {Tschernig}}, \bibinfo {author} {\bibfnamefont
  {A.}~\bibnamefont {Perez-Leija}}, \bibinfo {author} {\bibfnamefont
  {K.}~\bibnamefont {Busch}},\ and\ \bibinfo {author} {\bibfnamefont
  {R.~d.~J.}\ \bibnamefont {Le{\'o}n-Montiel}},\ }\bibfield  {title} {\bibinfo
  {title} {Topological protection in non-hermitian haldane honeycomb
  lattices},\ }\href@noop {} {\bibfield  {journal} {\bibinfo  {journal}
  {Physical Review Research}\ }\textbf {\bibinfo {volume} {2}},\ \bibinfo
  {pages} {013387} (\bibinfo {year} {2020})}\BibitemShut {NoStop}%
\bibitem [{\citenamefont {Zeuner}\ \emph {et~al.}(2015)\citenamefont {Zeuner},
  \citenamefont {Rechtsman}, \citenamefont {Plotnik}, \citenamefont {Lumer},
  \citenamefont {Nolte}, \citenamefont {Rudner}, \citenamefont {Segev},\ and\
  \citenamefont {Szameit}}]{zeuner2015observation}%
  \BibitemOpen
  \bibfield  {author} {\bibinfo {author} {\bibfnamefont {J.~M.}\ \bibnamefont
  {Zeuner}}, \bibinfo {author} {\bibfnamefont {M.~C.}\ \bibnamefont
  {Rechtsman}}, \bibinfo {author} {\bibfnamefont {Y.}~\bibnamefont {Plotnik}},
  \bibinfo {author} {\bibfnamefont {Y.}~\bibnamefont {Lumer}}, \bibinfo
  {author} {\bibfnamefont {S.}~\bibnamefont {Nolte}}, \bibinfo {author}
  {\bibfnamefont {M.~S.}\ \bibnamefont {Rudner}}, \bibinfo {author}
  {\bibfnamefont {M.}~\bibnamefont {Segev}},\ and\ \bibinfo {author}
  {\bibfnamefont {A.}~\bibnamefont {Szameit}},\ }\bibfield  {title} {\bibinfo
  {title} {Observation of a topological transition in the bulk of a
  non-hermitian system},\ }\href@noop {} {\bibfield  {journal} {\bibinfo
  {journal} {Physical review letters}\ }\textbf {\bibinfo {volume} {115}},\
  \bibinfo {pages} {040402} (\bibinfo {year} {2015})}\BibitemShut {NoStop}%
\bibitem [{\citenamefont {Weimann}\ \emph {et~al.}(2017)\citenamefont
  {Weimann}, \citenamefont {Kremer}, \citenamefont {Plotnik}, \citenamefont
  {Lumer}, \citenamefont {Nolte}, \citenamefont {Makris}, \citenamefont
  {Segev}, \citenamefont {Rechtsman},\ and\ \citenamefont
  {Szameit}}]{weimann2017topologically}%
  \BibitemOpen
  \bibfield  {author} {\bibinfo {author} {\bibfnamefont {S.}~\bibnamefont
  {Weimann}}, \bibinfo {author} {\bibfnamefont {M.}~\bibnamefont {Kremer}},
  \bibinfo {author} {\bibfnamefont {Y.}~\bibnamefont {Plotnik}}, \bibinfo
  {author} {\bibfnamefont {Y.}~\bibnamefont {Lumer}}, \bibinfo {author}
  {\bibfnamefont {S.}~\bibnamefont {Nolte}}, \bibinfo {author} {\bibfnamefont
  {K.~G.}\ \bibnamefont {Makris}}, \bibinfo {author} {\bibfnamefont
  {M.}~\bibnamefont {Segev}}, \bibinfo {author} {\bibfnamefont {M.~C.}\
  \bibnamefont {Rechtsman}},\ and\ \bibinfo {author} {\bibfnamefont
  {A.}~\bibnamefont {Szameit}},\ }\bibfield  {title} {\bibinfo {title}
  {Topologically protected bound states in photonic parity--time-symmetric
  crystals},\ }\href@noop {} {\bibfield  {journal} {\bibinfo  {journal} {Nature
  materials}\ }\textbf {\bibinfo {volume} {16}},\ \bibinfo {pages} {433}
  (\bibinfo {year} {2017})}\BibitemShut {NoStop}%
\bibitem [{\citenamefont {Liu}\ \emph {et~al.}(2021)\citenamefont {Liu},
  \citenamefont {Jung}, \citenamefont {Parto}, \citenamefont
  {Christodoulides},\ and\ \citenamefont {Khajavikhan}}]{liu2021gain}%
  \BibitemOpen
  \bibfield  {author} {\bibinfo {author} {\bibfnamefont {Y.~G.}\ \bibnamefont
  {Liu}}, \bibinfo {author} {\bibfnamefont {P.~S.}\ \bibnamefont {Jung}},
  \bibinfo {author} {\bibfnamefont {M.}~\bibnamefont {Parto}}, \bibinfo
  {author} {\bibfnamefont {D.~N.}\ \bibnamefont {Christodoulides}},\ and\
  \bibinfo {author} {\bibfnamefont {M.}~\bibnamefont {Khajavikhan}},\
  }\bibfield  {title} {\bibinfo {title} {Gain-induced topological response via
  tailored long-range interactions},\ }\href@noop {} {\bibfield  {journal}
  {\bibinfo  {journal} {Nature Physics}\ }\textbf {\bibinfo {volume} {17}},\
  \bibinfo {pages} {704} (\bibinfo {year} {2021})}\BibitemShut {NoStop}%
\bibitem [{\citenamefont {Weidemann}\ \emph {et~al.}(2022)\citenamefont
  {Weidemann}, \citenamefont {Kremer}, \citenamefont {Longhi},\ and\
  \citenamefont {Szameit}}]{weidemann2022topological}%
  \BibitemOpen
  \bibfield  {author} {\bibinfo {author} {\bibfnamefont {S.}~\bibnamefont
  {Weidemann}}, \bibinfo {author} {\bibfnamefont {M.}~\bibnamefont {Kremer}},
  \bibinfo {author} {\bibfnamefont {S.}~\bibnamefont {Longhi}},\ and\ \bibinfo
  {author} {\bibfnamefont {A.}~\bibnamefont {Szameit}},\ }\bibfield  {title}
  {\bibinfo {title} {Topological triple phase transition in non-hermitian
  floquet quasicrystals},\ }\href@noop {} {\bibfield  {journal} {\bibinfo
  {journal} {Nature}\ }\textbf {\bibinfo {volume} {601}},\ \bibinfo {pages}
  {354} (\bibinfo {year} {2022})}\BibitemShut {NoStop}%
\bibitem [{\citenamefont {Bahari}\ \emph {et~al.}(2017)\citenamefont {Bahari},
  \citenamefont {Ndao}, \citenamefont {Vallini}, \citenamefont {El~Amili},
  \citenamefont {Fainman},\ and\ \citenamefont
  {Kant{\'e}}}]{bahari2017nonreciprocal}%
  \BibitemOpen
  \bibfield  {author} {\bibinfo {author} {\bibfnamefont {B.}~\bibnamefont
  {Bahari}}, \bibinfo {author} {\bibfnamefont {A.}~\bibnamefont {Ndao}},
  \bibinfo {author} {\bibfnamefont {F.}~\bibnamefont {Vallini}}, \bibinfo
  {author} {\bibfnamefont {A.}~\bibnamefont {El~Amili}}, \bibinfo {author}
  {\bibfnamefont {Y.}~\bibnamefont {Fainman}},\ and\ \bibinfo {author}
  {\bibfnamefont {B.}~\bibnamefont {Kant{\'e}}},\ }\bibfield  {title} {\bibinfo
  {title} {Nonreciprocal lasing in topological cavities of arbitrary
  geometries},\ }\href@noop {} {\bibfield  {journal} {\bibinfo  {journal}
  {Science}\ }\textbf {\bibinfo {volume} {358}},\ \bibinfo {pages} {636}
  (\bibinfo {year} {2017})}\BibitemShut {NoStop}%
\bibitem [{\citenamefont {St-Jean}\ \emph {et~al.}(2017)\citenamefont
  {St-Jean}, \citenamefont {Goblot}, \citenamefont {Galopin}, \citenamefont
  {Lema{\^\i}tre}, \citenamefont {Ozawa}, \citenamefont {Le~Gratiet},
  \citenamefont {Sagnes}, \citenamefont {Bloch},\ and\ \citenamefont
  {Amo}}]{st2017lasing}%
  \BibitemOpen
  \bibfield  {author} {\bibinfo {author} {\bibfnamefont {P.}~\bibnamefont
  {St-Jean}}, \bibinfo {author} {\bibfnamefont {V.}~\bibnamefont {Goblot}},
  \bibinfo {author} {\bibfnamefont {E.}~\bibnamefont {Galopin}}, \bibinfo
  {author} {\bibfnamefont {A.}~\bibnamefont {Lema{\^\i}tre}}, \bibinfo {author}
  {\bibfnamefont {T.}~\bibnamefont {Ozawa}}, \bibinfo {author} {\bibfnamefont
  {L.}~\bibnamefont {Le~Gratiet}}, \bibinfo {author} {\bibfnamefont
  {I.}~\bibnamefont {Sagnes}}, \bibinfo {author} {\bibfnamefont
  {J.}~\bibnamefont {Bloch}},\ and\ \bibinfo {author} {\bibfnamefont
  {A.}~\bibnamefont {Amo}},\ }\bibfield  {title} {\bibinfo {title} {Lasing in
  topological edge states of a one-dimensional lattice},\ }\href@noop {}
  {\bibfield  {journal} {\bibinfo  {journal} {Nature Photonics}\ }\textbf
  {\bibinfo {volume} {11}},\ \bibinfo {pages} {651} (\bibinfo {year}
  {2017})}\BibitemShut {NoStop}%
\bibitem [{\citenamefont {Zhao}\ \emph {et~al.}(2018)\citenamefont {Zhao},
  \citenamefont {Miao}, \citenamefont {Teimourpour}, \citenamefont {Malzard},
  \citenamefont {El-Ganainy}, \citenamefont {Schomerus},\ and\ \citenamefont
  {Feng}}]{zhao2018topological}%
  \BibitemOpen
  \bibfield  {author} {\bibinfo {author} {\bibfnamefont {H.}~\bibnamefont
  {Zhao}}, \bibinfo {author} {\bibfnamefont {P.}~\bibnamefont {Miao}}, \bibinfo
  {author} {\bibfnamefont {M.~H.}\ \bibnamefont {Teimourpour}}, \bibinfo
  {author} {\bibfnamefont {S.}~\bibnamefont {Malzard}}, \bibinfo {author}
  {\bibfnamefont {R.}~\bibnamefont {El-Ganainy}}, \bibinfo {author}
  {\bibfnamefont {H.}~\bibnamefont {Schomerus}},\ and\ \bibinfo {author}
  {\bibfnamefont {L.}~\bibnamefont {Feng}},\ }\bibfield  {title} {\bibinfo
  {title} {Topological hybrid silicon microlasers},\ }\href@noop {} {\bibfield
  {journal} {\bibinfo  {journal} {Nature communications}\ }\textbf {\bibinfo
  {volume} {9}},\ \bibinfo {pages} {981} (\bibinfo {year} {2018})}\BibitemShut
  {NoStop}%
\bibitem [{\citenamefont {Takata}\ and\ \citenamefont
  {Notomi}(2018)}]{TakataPRL2018}%
  \BibitemOpen
  \bibfield  {author} {\bibinfo {author} {\bibfnamefont {K.}~\bibnamefont
  {Takata}}\ and\ \bibinfo {author} {\bibfnamefont {M.}~\bibnamefont
  {Notomi}},\ }\bibfield  {title} {\bibinfo {title} {Photonic topological
  insulating phase induced solely by gain and loss},\ }\href
  {https://doi.org/10.1103/PhysRevLett.121.213902} {\bibfield  {journal}
  {\bibinfo  {journal} {Phys. Rev. Lett.}\ }\textbf {\bibinfo {volume} {121}},\
  \bibinfo {pages} {213902} (\bibinfo {year} {2018})}\BibitemShut {NoStop}%
\bibitem [{\citenamefont {Zhu}\ \emph {et~al.}(2023)\citenamefont {Zhu},
  \citenamefont {Lang}, \citenamefont {Wang}, \citenamefont {Wang},\ and\
  \citenamefont {Chong}}]{BofengPRR2023}%
  \BibitemOpen
  \bibfield  {author} {\bibinfo {author} {\bibfnamefont {B.}~\bibnamefont
  {Zhu}}, \bibinfo {author} {\bibfnamefont {L.-J.}\ \bibnamefont {Lang}},
  \bibinfo {author} {\bibfnamefont {Q.}~\bibnamefont {Wang}}, \bibinfo {author}
  {\bibfnamefont {Q.~J.}\ \bibnamefont {Wang}},\ and\ \bibinfo {author}
  {\bibfnamefont {Y.~D.}\ \bibnamefont {Chong}},\ }\bibfield  {title} {\bibinfo
  {title} {Topological transitions with an imaginary aubry-andr\'e-harper
  potential},\ }\href {https://doi.org/10.1103/PhysRevResearch.5.023044}
  {\bibfield  {journal} {\bibinfo  {journal} {Phys. Rev. Res.}\ }\textbf
  {\bibinfo {volume} {5}},\ \bibinfo {pages} {023044} (\bibinfo {year}
  {2023})}\BibitemShut {NoStop}%
\bibitem [{\citenamefont {Pereira}\ \emph {et~al.}(2024)\citenamefont
  {Pereira}, \citenamefont {Li}, \citenamefont {Blanco-Redondo},\ and\
  \citenamefont {Lado}}]{JosePPR2024}%
  \BibitemOpen
  \bibfield  {author} {\bibinfo {author} {\bibfnamefont {E.~L.}\ \bibnamefont
  {Pereira}}, \bibinfo {author} {\bibfnamefont {H.}~\bibnamefont {Li}},
  \bibinfo {author} {\bibfnamefont {A.}~\bibnamefont {Blanco-Redondo}},\ and\
  \bibinfo {author} {\bibfnamefont {J.~L.}\ \bibnamefont {Lado}},\ }\bibfield
  {title} {\bibinfo {title} {Non-hermitian topology and criticality in photonic
  arrays with engineered losses},\ }\href
  {https://doi.org/10.1103/PhysRevResearch.6.023004} {\bibfield  {journal}
  {\bibinfo  {journal} {Phys. Rev. Res.}\ }\textbf {\bibinfo {volume} {6}},\
  \bibinfo {pages} {023004} (\bibinfo {year} {2024})}\BibitemShut {NoStop}%
\bibitem [{\citenamefont {Liu}\ \emph {et~al.}(2020)\citenamefont {Liu},
  \citenamefont {Ma}, \citenamefont {Yang}, \citenamefont {Zhang},
  \citenamefont {Gao}, \citenamefont {Xiang}, \citenamefont {Cui},\ and\
  \citenamefont {Zhang}}]{Electrical_PRApp2020}%
  \BibitemOpen
  \bibfield  {author} {\bibinfo {author} {\bibfnamefont {S.}~\bibnamefont
  {Liu}}, \bibinfo {author} {\bibfnamefont {S.}~\bibnamefont {Ma}}, \bibinfo
  {author} {\bibfnamefont {C.}~\bibnamefont {Yang}}, \bibinfo {author}
  {\bibfnamefont {L.}~\bibnamefont {Zhang}}, \bibinfo {author} {\bibfnamefont
  {W.}~\bibnamefont {Gao}}, \bibinfo {author} {\bibfnamefont {Y.~J.}\
  \bibnamefont {Xiang}}, \bibinfo {author} {\bibfnamefont {T.~J.}\ \bibnamefont
  {Cui}},\ and\ \bibinfo {author} {\bibfnamefont {S.}~\bibnamefont {Zhang}},\
  }\bibfield  {title} {\bibinfo {title} {Gain- and loss-induced topological
  insulating phase in a non-hermitian electrical circuit},\ }\href
  {https://doi.org/10.1103/PhysRevApplied.13.014047} {\bibfield  {journal}
  {\bibinfo  {journal} {Phys. Rev. Appl.}\ }\textbf {\bibinfo {volume} {13}},\
  \bibinfo {pages} {014047} (\bibinfo {year} {2020})}\BibitemShut {NoStop}%
\bibitem [{\citenamefont {Gao}\ \emph {et~al.}(2020)\citenamefont {Gao},
  \citenamefont {Xue}, \citenamefont {Wang}, \citenamefont {Gu}, \citenamefont
  {Liu}, \citenamefont {Zhu},\ and\ \citenamefont {Zhang}}]{Acoustic_PRB2020}%
  \BibitemOpen
  \bibfield  {author} {\bibinfo {author} {\bibfnamefont {H.}~\bibnamefont
  {Gao}}, \bibinfo {author} {\bibfnamefont {H.}~\bibnamefont {Xue}}, \bibinfo
  {author} {\bibfnamefont {Q.}~\bibnamefont {Wang}}, \bibinfo {author}
  {\bibfnamefont {Z.}~\bibnamefont {Gu}}, \bibinfo {author} {\bibfnamefont
  {T.}~\bibnamefont {Liu}}, \bibinfo {author} {\bibfnamefont {J.}~\bibnamefont
  {Zhu}},\ and\ \bibinfo {author} {\bibfnamefont {B.}~\bibnamefont {Zhang}},\
  }\bibfield  {title} {\bibinfo {title} {Observation of topological edge states
  induced solely by non-hermiticity in an acoustic crystal},\ }\href
  {https://doi.org/10.1103/PhysRevB.101.180303} {\bibfield  {journal} {\bibinfo
   {journal} {Phys. Rev. B}\ }\textbf {\bibinfo {volume} {101}},\ \bibinfo
  {pages} {180303} (\bibinfo {year} {2020})}\BibitemShut {NoStop}%
\bibitem [{\citenamefont {Gao}\ \emph {et~al.}(2021)\citenamefont {Gao},
  \citenamefont {Xue}, \citenamefont {Gu}, \citenamefont {Liu}, \citenamefont
  {Zhu},\ and\ \citenamefont {Zhang}}]{gao2021non}%
  \BibitemOpen
  \bibfield  {author} {\bibinfo {author} {\bibfnamefont {H.}~\bibnamefont
  {Gao}}, \bibinfo {author} {\bibfnamefont {H.}~\bibnamefont {Xue}}, \bibinfo
  {author} {\bibfnamefont {Z.}~\bibnamefont {Gu}}, \bibinfo {author}
  {\bibfnamefont {T.}~\bibnamefont {Liu}}, \bibinfo {author} {\bibfnamefont
  {J.}~\bibnamefont {Zhu}},\ and\ \bibinfo {author} {\bibfnamefont
  {B.}~\bibnamefont {Zhang}},\ }\bibfield  {title} {\bibinfo {title}
  {Non-hermitian route to higher-order topology in an acoustic crystal},\
  }\href@noop {} {\bibfield  {journal} {\bibinfo  {journal} {Nature
  communications}\ }\textbf {\bibinfo {volume} {12}},\ \bibinfo {pages} {1888}
  (\bibinfo {year} {2021})}\BibitemShut {NoStop}%
\bibitem [{\citenamefont {Fan}\ \emph {et~al.}(2022)\citenamefont {Fan},
  \citenamefont {Gao}, \citenamefont {An}, \citenamefont {Gu}, \citenamefont
  {Liang}, \citenamefont {Zheng},\ and\ \citenamefont
  {Liu}}]{Elastic_MechSys2022}%
  \BibitemOpen
  \bibfield  {author} {\bibinfo {author} {\bibfnamefont {H.}~\bibnamefont
  {Fan}}, \bibinfo {author} {\bibfnamefont {H.}~\bibnamefont {Gao}}, \bibinfo
  {author} {\bibfnamefont {S.}~\bibnamefont {An}}, \bibinfo {author}
  {\bibfnamefont {Z.}~\bibnamefont {Gu}}, \bibinfo {author} {\bibfnamefont
  {S.}~\bibnamefont {Liang}}, \bibinfo {author} {\bibfnamefont
  {Y.}~\bibnamefont {Zheng}},\ and\ \bibinfo {author} {\bibfnamefont
  {T.}~\bibnamefont {Liu}},\ }\bibfield  {title} {\bibinfo {title} {Hermitian
  and non-hermitian topological edge states in one-dimensional perturbative
  elastic metamaterials},\ }\href
  {https://doi.org/https://doi.org/10.1016/j.ymssp.2021.108774} {\bibfield
  {journal} {\bibinfo  {journal} {Mechanical Systems and Signal Processing}\
  }\textbf {\bibinfo {volume} {169}},\ \bibinfo {pages} {108774} (\bibinfo
  {year} {2022})}\BibitemShut {NoStop}%
\bibitem [{\citenamefont {Wetter}\ \emph {et~al.}(2023)\citenamefont {Wetter},
  \citenamefont {Fleischhauer}, \citenamefont {Linden},\ and\ \citenamefont
  {Schmitt}}]{Plasmonic_PRL2023}%
  \BibitemOpen
  \bibfield  {author} {\bibinfo {author} {\bibfnamefont {H.}~\bibnamefont
  {Wetter}}, \bibinfo {author} {\bibfnamefont {M.}~\bibnamefont
  {Fleischhauer}}, \bibinfo {author} {\bibfnamefont {S.}~\bibnamefont
  {Linden}},\ and\ \bibinfo {author} {\bibfnamefont {J.}~\bibnamefont
  {Schmitt}},\ }\bibfield  {title} {\bibinfo {title} {Observation of a
  topological edge state stabilized by dissipation},\ }\href
  {https://doi.org/10.1103/PhysRevLett.131.083801} {\bibfield  {journal}
  {\bibinfo  {journal} {Phys. Rev. Lett.}\ }\textbf {\bibinfo {volume} {131}},\
  \bibinfo {pages} {083801} (\bibinfo {year} {2023})}\BibitemShut {NoStop}%
\bibitem [{\citenamefont {On}\ \emph {et~al.}(2024)\citenamefont {On},
  \citenamefont {Ashtiani}, \citenamefont {Sanchez-Jacome}, \citenamefont
  {Perez-Lopez}, \citenamefont {Yoo},\ and\ \citenamefont
  {Blanco-Redondo}}]{MehmetNatComm2024}%
  \BibitemOpen
  \bibfield  {author} {\bibinfo {author} {\bibfnamefont {M.~B.}\ \bibnamefont
  {On}}, \bibinfo {author} {\bibfnamefont {F.}~\bibnamefont {Ashtiani}},
  \bibinfo {author} {\bibfnamefont {D.}~\bibnamefont {Sanchez-Jacome}},
  \bibinfo {author} {\bibfnamefont {D.}~\bibnamefont {Perez-Lopez}}, \bibinfo
  {author} {\bibfnamefont {S.~J.~B.}\ \bibnamefont {Yoo}},\ and\ \bibinfo
  {author} {\bibfnamefont {A.}~\bibnamefont {Blanco-Redondo}},\ }\bibfield
  {title} {\bibinfo {title} {Programmable integrated photonics for topological
  hamiltonians},\ }\href {https://doi.org/10.1038/s41467-024-44939-3}
  {\bibfield  {journal} {\bibinfo  {journal} {Nature Communications}\ }\textbf
  {\bibinfo {volume} {15}},\ \bibinfo {pages} {629} (\bibinfo {year}
  {2024})}\BibitemShut {NoStop}%
\bibitem [{\citenamefont {Aubry}\ and\ \citenamefont
  {Andre}(1980)}]{AubryAndre1980}%
  \BibitemOpen
  \bibfield  {author} {\bibinfo {author} {\bibfnamefont {S.}~\bibnamefont
  {Aubry}}\ and\ \bibinfo {author} {\bibfnamefont {G.}~\bibnamefont {Andre}},\
  }\bibfield  {title} {\bibinfo {title} {Analyticity breaking and anderson
  localization in incommensurate lattices},\ }\href@noop {} {\bibfield
  {journal} {\bibinfo  {journal} {Ann. Isr. Phys. Soc.}\ }\textbf {\bibinfo
  {volume} {3}},\ \bibinfo {pages} {133} (\bibinfo {year} {1980})}\BibitemShut
  {NoStop}%
\bibitem [{\citenamefont {Verbin}\ \emph {et~al.}(2015)\citenamefont {Verbin},
  \citenamefont {Zilberberg}, \citenamefont {Lahini}, \citenamefont {Kraus},\
  and\ \citenamefont {Silberberg}}]{ZilberbergPRB2015}%
  \BibitemOpen
  \bibfield  {author} {\bibinfo {author} {\bibfnamefont {M.}~\bibnamefont
  {Verbin}}, \bibinfo {author} {\bibfnamefont {O.}~\bibnamefont {Zilberberg}},
  \bibinfo {author} {\bibfnamefont {Y.}~\bibnamefont {Lahini}}, \bibinfo
  {author} {\bibfnamefont {Y.~E.}\ \bibnamefont {Kraus}},\ and\ \bibinfo
  {author} {\bibfnamefont {Y.}~\bibnamefont {Silberberg}},\ }\bibfield  {title}
  {\bibinfo {title} {Topological pumping over a photonic fibonacci
  quasicrystal},\ }\href {https://doi.org/10.1103/PhysRevB.91.064201}
  {\bibfield  {journal} {\bibinfo  {journal} {Phys. Rev. B}\ }\textbf {\bibinfo
  {volume} {91}},\ \bibinfo {pages} {064201} (\bibinfo {year}
  {2015})}\BibitemShut {NoStop}%
\bibitem [{\citenamefont {Tambasco}\ \emph {et~al.}(2018)\citenamefont
  {Tambasco}, \citenamefont {Corrielli}, \citenamefont {Chapman}, \citenamefont
  {Crespi}, \citenamefont {Zilberberg}, \citenamefont {Osellame},\ and\
  \citenamefont {Peruzzo}}]{TambascoScience2018}%
  \BibitemOpen
  \bibfield  {author} {\bibinfo {author} {\bibfnamefont {J.-L.}\ \bibnamefont
  {Tambasco}}, \bibinfo {author} {\bibfnamefont {G.}~\bibnamefont {Corrielli}},
  \bibinfo {author} {\bibfnamefont {R.~J.}\ \bibnamefont {Chapman}}, \bibinfo
  {author} {\bibfnamefont {A.}~\bibnamefont {Crespi}}, \bibinfo {author}
  {\bibfnamefont {O.}~\bibnamefont {Zilberberg}}, \bibinfo {author}
  {\bibfnamefont {R.}~\bibnamefont {Osellame}},\ and\ \bibinfo {author}
  {\bibfnamefont {A.}~\bibnamefont {Peruzzo}},\ }\bibfield  {title} {\bibinfo
  {title} {Quantum interference of topological states of light},\ }\href
  {https://doi.org/10.1126/sciadv.aat3187} {\bibfield  {journal} {\bibinfo
  {journal} {Science Advances}\ }\textbf {\bibinfo {volume} {4}},\ \bibinfo
  {pages} {eaat3187} (\bibinfo {year} {2018})},\ \Eprint
  {https://arxiv.org/abs/https://www.science.org/doi/pdf/10.1126/sciadv.aat3187}
  {https://www.science.org/doi/pdf/10.1126/sciadv.aat3187} \BibitemShut
  {NoStop}%
\bibitem [{\citenamefont {Yuce}(2014)}]{YucePLA2014}%
  \BibitemOpen
  \bibfield  {author} {\bibinfo {author} {\bibfnamefont {C.}~\bibnamefont
  {Yuce}},\ }\bibfield  {title} {\bibinfo {title} {Pt symmetric aubry–andre
  model},\ }\href
  {https://doi.org/https://doi.org/10.1016/j.physleta.2014.05.005} {\bibfield
  {journal} {\bibinfo  {journal} {Phys. Lett. A}\ }\textbf {\bibinfo {volume}
  {378}},\ \bibinfo {pages} {2024} (\bibinfo {year} {2014})}\BibitemShut
  {NoStop}%
\bibitem [{\citenamefont {Longhi}(2019)}]{LonghiPRB2019}%
  \BibitemOpen
  \bibfield  {author} {\bibinfo {author} {\bibfnamefont {S.}~\bibnamefont
  {Longhi}},\ }\bibfield  {title} {\bibinfo {title} {Metal-insulator phase
  transition in a non-hermitian aubry-andr\'e-harper model},\ }\href
  {https://doi.org/10.1103/PhysRevB.100.125157} {\bibfield  {journal} {\bibinfo
   {journal} {Phys. Rev. B}\ }\textbf {\bibinfo {volume} {100}},\ \bibinfo
  {pages} {125157} (\bibinfo {year} {2019})}\BibitemShut {NoStop}%
\bibitem [{\citenamefont {Zeng}\ \emph {et~al.}(2020)\citenamefont {Zeng},
  \citenamefont {Yang},\ and\ \citenamefont {Xu}}]{ZengPRB2020}%
  \BibitemOpen
  \bibfield  {author} {\bibinfo {author} {\bibfnamefont {Q.-B.}\ \bibnamefont
  {Zeng}}, \bibinfo {author} {\bibfnamefont {Y.-B.}\ \bibnamefont {Yang}},\
  and\ \bibinfo {author} {\bibfnamefont {Y.}~\bibnamefont {Xu}},\ }\bibfield
  {title} {\bibinfo {title} {Topological phases in non-hermitian
  aubry-andr\'e-harper models},\ }\href
  {https://doi.org/10.1103/PhysRevB.101.020201} {\bibfield  {journal} {\bibinfo
   {journal} {Phys. Rev. B}\ }\textbf {\bibinfo {volume} {101}},\ \bibinfo
  {pages} {020201} (\bibinfo {year} {2020})}\BibitemShut {NoStop}%
\end{thebibliography}%


\begin{thebibliography}{5}%
\makeatletter
\providecommand \@ifxundefined [1]{%
 \@ifx{#1\undefined}
}%
\providecommand \@ifnum [1]{%
 \ifnum #1\expandafter \@firstoftwo
 \else \expandafter \@secondoftwo
 \fi
}%
\providecommand \@ifx [1]{%
 \ifx #1\expandafter \@firstoftwo
 \else \expandafter \@secondoftwo
 \fi
}%
\providecommand \natexlab [1]{#1}%
\providecommand \enquote  [1]{``#1''}%
\providecommand \bibnamefont  [1]{#1}%
\providecommand \bibfnamefont [1]{#1}%
\providecommand \citenamefont [1]{#1}%
\providecommand \href@noop [0]{\@secondoftwo}%
\providecommand \href [0]{\begingroup \@sanitize@url \@href}%
\providecommand \@href[1]{\@@startlink{#1}\@@href}%
\providecommand \@@href[1]{\endgroup#1\@@endlink}%
\providecommand \@sanitize@url [0]{\catcode `\\12\catcode `\$12\catcode
  `\&12\catcode `\#12\catcode `\^12\catcode `\_12\catcode `\%12\relax}%
\providecommand \@@startlink[1]{}%
\providecommand \@@endlink[0]{}%
\providecommand \url  [0]{\begingroup\@sanitize@url \@url }%
\providecommand \@url [1]{\endgroup\@href {#1}{\urlprefix }}%
\providecommand \urlprefix  [0]{URL }%
\providecommand \Eprint [0]{\href }%
\providecommand \doibase [0]{https://doi.org/}%
\providecommand \selectlanguage [0]{\@gobble}%
\providecommand \bibinfo  [0]{\@secondoftwo}%
\providecommand \bibfield  [0]{\@secondoftwo}%
\providecommand \translation [1]{[#1]}%
\providecommand \BibitemOpen [0]{}%
\providecommand \bibitemStop [0]{}%
\providecommand \bibitemNoStop [0]{.\EOS\space}%
\providecommand \EOS [0]{\spacefactor3000\relax}%
\providecommand \BibitemShut  [1]{\csname bibitem#1\endcsname}%
\let\auto@bib@innerbib\@empty
\bibitem [{\citenamefont {Van}(2016)}]{OpticalMicroRingbook}%
  \BibitemOpen
  \bibfield  {author} {\bibinfo {author} {\bibfnamefont {V.}~\bibnamefont
  {Van}},\ }\href@noop {} {\emph {\bibinfo {title} {Optical Microring
  Resonators: Theory, Techniques, and Applications}}}\ (\bibinfo  {publisher}
  {CRC Press},\ \bibinfo {year} {2016})\BibitemShut {NoStop}%
\bibitem [{\citenamefont {Hashemi}\ \emph
  {et~al.}(2022{\natexlab{a}})\citenamefont {Hashemi}, \citenamefont {Busch},
  \citenamefont {Ozdemir},\ and\ \citenamefont {El-Ganainy}}]{HashemiPRR2022}%
  \BibitemOpen
  \bibfield  {author} {\bibinfo {author} {\bibfnamefont {A.}~\bibnamefont
  {Hashemi}}, \bibinfo {author} {\bibfnamefont {K.}~\bibnamefont {Busch}},
  \bibinfo {author} {\bibfnamefont {S.~K.}\ \bibnamefont {Ozdemir}},\ and\
  \bibinfo {author} {\bibfnamefont {R.}~\bibnamefont {El-Ganainy}},\ }\bibfield
   {title} {\bibinfo {title} {Uniform optical gain as a non-hermitian control
  knob},\ }\href {https://doi.org/10.1103/PhysRevResearch.4.043169} {\bibfield
  {journal} {\bibinfo  {journal} {Phys. Rev. Res.}\ }\textbf {\bibinfo {volume}
  {4}},\ \bibinfo {pages} {043169} (\bibinfo {year}
  {2022}{\natexlab{a}})}\BibitemShut {NoStop}%
\bibitem [{\citenamefont {Suh}\ \emph {et~al.}(2004)\citenamefont {Suh},
  \citenamefont {Wang},\ and\ \citenamefont {Fan}}]{WonjooJQE2004}%
  \BibitemOpen
  \bibfield  {author} {\bibinfo {author} {\bibfnamefont {W.}~\bibnamefont
  {Suh}}, \bibinfo {author} {\bibfnamefont {Z.}~\bibnamefont {Wang}},\ and\
  \bibinfo {author} {\bibfnamefont {S.}~\bibnamefont {Fan}},\ }\bibfield
  {title} {\bibinfo {title} {Temporal coupled-mode theory and the presence of
  non-orthogonal modes in lossless multimode cavities},\ }\href
  {https://doi.org/10.1109/JQE.2004.834773} {\bibfield  {journal} {\bibinfo
  {journal} {IEEE Journal of Quantum Electronics}\ }\textbf {\bibinfo {volume}
  {40}},\ \bibinfo {pages} {1511} (\bibinfo {year} {2004})}\BibitemShut
  {NoStop}%
\bibitem [{\citenamefont {Economou}(2010)}]{greenfunctionbook}%
  \BibitemOpen
  \bibfield  {author} {\bibinfo {author} {\bibfnamefont {E.~N.}\ \bibnamefont
  {Economou}},\ }\href@noop {} {\emph {\bibinfo {title} {Green's Functions in
  Quantum Physics}}}\ (\bibinfo  {publisher} {Springer Berlin Heidelberg},\
  \bibinfo {year} {2010})\BibitemShut {NoStop}%
\bibitem [{\citenamefont {Hashemi}\ \emph
  {et~al.}(2022{\natexlab{b}})\citenamefont {Hashemi}, \citenamefont {Busch},
  \citenamefont {Christodoulides}, \citenamefont {Ozdemir},\ and\ \citenamefont
  {El-Ganainy}}]{HashemiNatComm2022}%
  \BibitemOpen
  \bibfield  {author} {\bibinfo {author} {\bibfnamefont {A.}~\bibnamefont
  {Hashemi}}, \bibinfo {author} {\bibfnamefont {K.}~\bibnamefont {Busch}},
  \bibinfo {author} {\bibfnamefont {D.~N.}\ \bibnamefont {Christodoulides}},
  \bibinfo {author} {\bibfnamefont {S.~K.}\ \bibnamefont {Ozdemir}},\ and\
  \bibinfo {author} {\bibfnamefont {R.}~\bibnamefont {El-Ganainy}},\ }\bibfield
   {title} {\bibinfo {title} {Linear response theory of open systems with
  exceptional points},\ }\href {https://doi.org/10.1038/s41467-022-30715-8}
  {\bibfield  {journal} {\bibinfo  {journal} {Nature Communications}\ }\textbf
  {\bibinfo {volume} {13}},\ \bibinfo {pages} {3281} (\bibinfo {year}
  {2022}{\natexlab{b}})}\BibitemShut {NoStop}%
\end{thebibliography}%

\section*{Acknowledgments}
A.B.-R. acknowledges support by the NSF award no. 2328993. E.L.P. acknowledges support from the Nokia Industrial Doctoral School in Quantum Technology. J.L.L. acknowledges the financial support from the Academy of Finland Projects No. 331342 and No. 358088 and the Jane and Aatos Erkko Foundation. E.L.P. and J.L.L. acknowledge the computational resources provided by the Aalto Science-IT project.

\end{document}


\begin{center}
\textbf{{\Large Supplementary Information:}}\\[0.6cm]

\textbf{{\Large Observation of non-Hermitian topology from optical loss modulation}}\\[0.6cm]

{Amin Hashemi, Elizabeth Louis Pereira, Hongwei Li, Jose L. Lado, Andrea Blanco-Redondo}
\end{center}

\newpage

\section*{Supplementary notes}
\subsection{\textit{Programmable silicon photonics mesh}}
\begin{figure}[b!]
    \centering
    \includegraphics[width=4.2in]{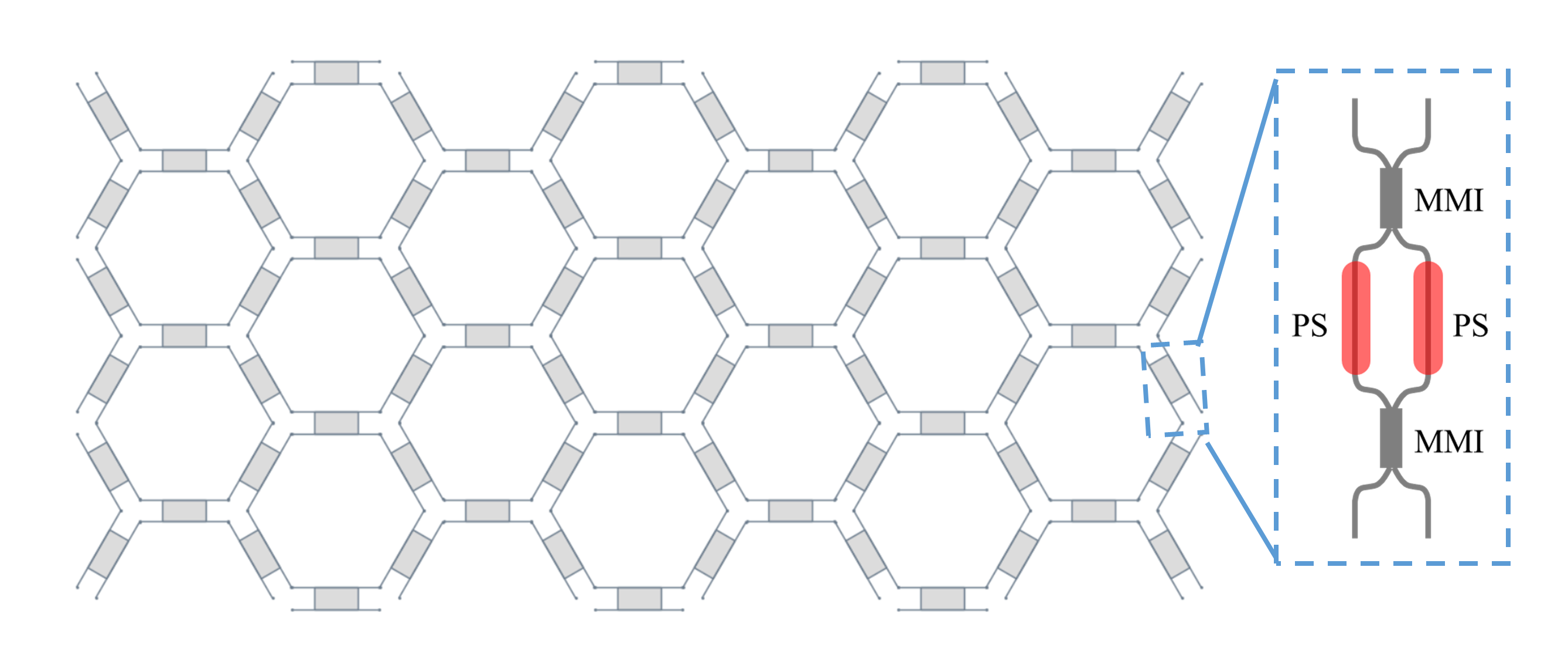}
    \caption{\textbf{The programmable integrated photonic platform.} The schematic of the programmable integrated photonic platform composed of the hexagonal mesh of Mach-Zehnder interferometers (MZIs). Inset shows one MZI consisting of two Multi-mode interferometers (MMIs) and phase shifters.}
    \label{fig:mesh_ipronics}
\end{figure}
The programmable integrated photonics platform used for the experimental implementation of the models consists of a hexagonal mesh of programmable silicon photonics Mach–Zehnder interferometers (MZIs), as shown in Supplementary Fig. \ref{fig:mesh_ipronics}. Each MZI comprises two \(2 \times 2\) multi-mode interferometers (MMIs) and two phase shifters (PSs), detailed in the inset of Supplementary Fig. \ref{fig:mesh_ipronics}. The length of each MZI is 811 $\mu$m, with an average insertion loss of 0.5 dB. The MZIs can be reconfigured by adjusting the voltage applied to the PSs, allowing precise control over the power splitting and the relative phase of the fields at the two outputs of each MZI. The platform includes 72 reconfigurable MZIs and 28 optical input/output ports located at the edges of the mesh for coupling light in and out of the chip. The output of a tunable laser is coupled to the mesh with about 2–4 dB fiber to chip coupling loss and an additional 6–8 dB on-chip routing loss from the input coupler to the desired MZI. The laser is thermally tunable from 1549.9 nm to 1550.2 nm with a resolution of 3 pm. The power measurements shown in this manuscript are taken with integrated photodetectors (PDs) with a -70 dBm sensitivity. All input/output ports, MZIs, and PDs are configurable and controllable via a Python-based software development kit (SDK). Optical, electronic, and software functionality is contained into a plug-and-play device (\textit{iPronics Smartlight Processor}).

\subsection{\textit{Experimental characterization of the model parameters}}
\begin{figure*}[b!]
    \centering
    \includegraphics[width=\textwidth]{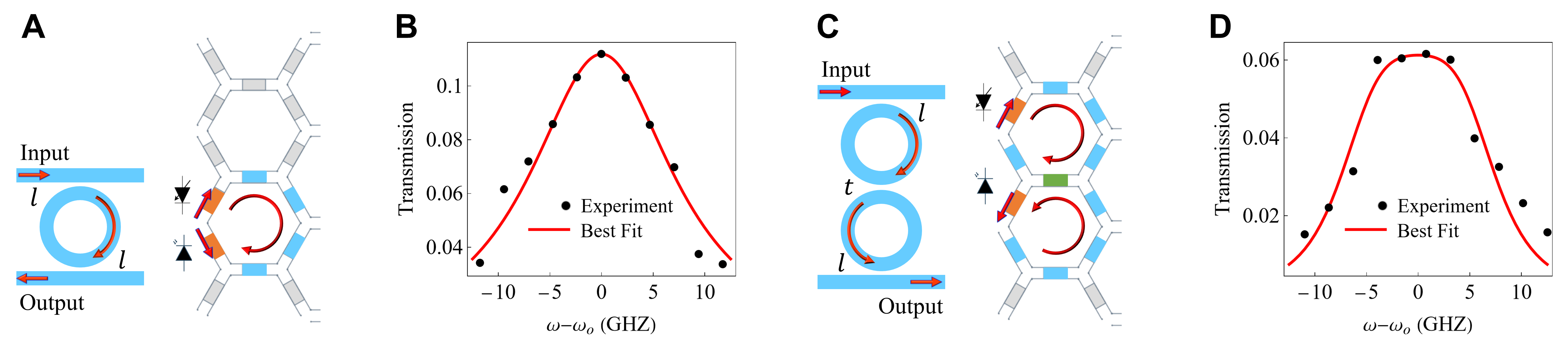}
    \caption{\textbf{The coupling factors characterization. A}, The schematic of the add-drop filter with a single ring resonator and its implementation on the programmable platform. \textbf{B}, Plot of the transmission power of the add-drop filter with a single ring resonator as a function of excitation frequency. \textbf{C}, The schematic of the add-drop filter with two ring resonators and its implementation on the programmable chip. \textbf{D}, Plot of the transmission power of the add-drop filter with two ring resonators against excitation frequency.}
    \label{fig:ipronics_couplings}
\end{figure*}
Ring resonators are the building block of all of our experimental model implementations. By configuring the power splitting of the MZIs at the edge of a hexagon, we can trap the light within the hexagon, thereby generating a resonator akin to a micro ring resonator. It is also feasible to couple input/output ports to a ring resonator by adjusting the power splitting factor of some of the ring resonator's MZIs (shown with brown color in Supplementary Fig. \ref{fig:ipronics_couplings}A).\\ 
To measure the field decay rate, $l$, between the ring resonator and an input/output port, we create an add-drop ring resonator filter, as depicted in Supplementary Fig. \ref{fig:ipronics_couplings}A. By fitting the transmission power spectrum measured from the output port with the Lorentzian distribution \cite{OpticalMicroRingbook}:
\begin{equation}
    \frac{p_{drop}}{p_{in}}=\frac{4\l^2}{\Delta^2+(2l+l_o)^2},
\end{equation}
the field decay rate to the waveguides, $l$ can be determined. Here $\Delta=\omega-\omega_o$, where $\omega_o$ is the resonance frequency of the micro ring resonator, and $l_o$ denotes the intrinsic field dacay rate of the resonator. For instance, Supplementary Fig. \ref{fig:ipronics_couplings}B, represents the experimental data (black dots) and the fitted curve (red curve), resulting in $l=0.23$ GHz for the configuration of the add-drop filter with the power splitting of $k_l=0.2$ for MZIs at input/output ports (shown in brown color in Supplementary Fig. \ref{fig:ipronics_couplings}A).\\
To evaluate the coupling between two ring resonators, $t$, we consider the add-drop filter with two ring resonators as illustrated in Supplementary Fig. \ref{fig:ipronics_couplings}C. In this case the transmission power from output port is given by \cite{HashemiPRR2022}:
\begin{equation}
    \frac{p_{drop}}{p_{in}}=\frac{4\l^2 t^2}{\Delta^4+2\Delta^2\left[(l+l_o)^2-t^2\right]+\left[(l+l_o)^2+t^2\right]^2}.
    \label{eq:2ring_T}
\end{equation}
Supplementary Fig. \ref{fig:ipronics_couplings}D presents the experimental data for the power transmission of the add-drop filter with two rings, depicted by black points; the red solid line represents the curve obtained by fitting this data using Eq.(\ref{eq:2ring_T}). In this setup, the power splitting factor $k_t=0.15$ for MZI common to both rings (shown in green in Supplementary Fig. \ref{fig:ipronics_couplings}C) results in a transmission coefficient $t=0.82$ GHz.

\subsection{\textit{Calculation of the localization degree of the states}}
To quantify the localization of the eigenmodes we calculate a quantity we call the localization degree ($LD$). The LD of a state, $\ket{\psi}$, is calculated using the following expression:
\begin{equation}
    LD={|\braket{\psi|P_1|\psi}|^2+|\braket{\psi|P_N|\psi}|^2},
\end{equation}
where $P_{1,N}$ are projection operators, projecting on the first and last site of the lattice, respectively. States with a large (small) LD are strongly localized (delocalized). Consequently, the LD quantity assists us in distinguishing edge and bulk modes (see Figs.2C in the main text).
The localization of states in the right or left edge can be extracted by computing the following expectation value:
\begin{equation}
    LD1={-|\braket{\psi|P_1|\psi}|^2+|\braket{\psi|P_N|\psi}|^2}.
\end{equation}
In this case states which are localized at the left or right edge of the lattice will correspondingly possess a negative or positive value of $LD1$ (see Figs.3C in the main text)
\subsection{\textit{Non-Hermitian systems excitation}}
Using TCMT \cite{WonjooJQE2004}, a linear resonant photonics structure can be described by the following equation:
\begin{equation}
    i\frac{d\ket{a(t)}}{dt}=H\ket{a(t)}+i\Gamma\ket{b(t)},
    \label{eq:TCMT}
\end{equation}
where, $\ket{a(t)}=[a_1(t),a_2(t),\cdots,a_N(t)]^T$ and $\ket{b(t)}=[b_1(t),b_2(t),\cdots,b_L(t)]^T$ represent the field amplitudes of the resonant modes and the input/output channels, respectively. The $N\times N$ coupling matrix $H$ (also called the Hamiltonian) includes the resonant frequencies and the coupling coefficients, while the $N\times L$ matrix $\Gamma$ describes the coupling coefficients between resonant modes and input/output channels.\\
In a closed (Hermitian) system, where no energy exchange occurs with the environment, the system's Hamiltonian is a Hermitian matrix. Conversely, in an open (non-Hermitian) system that permits energy exchange with the environment, the system's Hamiltonian is a non-Hermitian matrix. The steady state response of a system for an excitation is given by the system's resolvent (also called the Green's operator) \cite{greenfunctionbook}, $G(\omega)\equiv \left(\omega I - H\right)^{-1}$, as:
\begin{equation}
    \ket{A(\omega)}=G(\omega)\ket{S(\omega)},
\end{equation}
where $\ket{A(\omega)}\equiv \mathcal{FT}\left(\ket{a(t)}\right)$, $\ket{S(\omega)}\equiv \mathcal{FT}\left(i\Gamma\ket{b(t)}\right)$, and $\mathcal{FT}$ stands for the Fourier transform.
In a Hermitian system the resolvent is expressed in terms of the Hamiltonian eigenvectors, $\ket{\psi_n}$, and eigenvalues, $\Omega_n$, as:
\begin{equation}
    G(\omega) = \sum_{n=1}^{N} \frac{\ket{\psi_n}\bra{\psi_n}}{\omega-\Omega_n}.
    \label{eq:resolvent_Hermitian}
\end{equation}
Considering that in a Hermitian system, all eigenvalues, $\Omega_n$, are real numbers, thus, if the excitation frequency of the input signal aligns with one of the system's eigenfrequencies (say $\Omega_i$), even a weak non-zero overlap of the input signal with the corresponding eigenmode, ($\ket{\psi_i}$), causes the term proportional to $1/(\omega-\Omega_i)$ in the Green's operator expansion to diverge. This implies that other terms in the Green's operator expansion are negligible compared to this term, meaning that the excitation signal is fully transferred to the corresponding eigenmode. However, in reality, nonlinear effects modulate the dynamics, preventing the infinite growth of the mode amplitudes. Hence, to excite one of the Eigenmodes in a Hermitian system, adjusting the excitation frequency to match the corresponding eigenfrequency and ensuring spatial overlap is sufficient.\\
In non-Hermitian systems --assuming that the system does not have any exceptional point -- the resolvent is expressed by \cite{HashemiNatComm2022}:
\begin{equation}
    G(\omega)=\sum_{n=1}^N\frac{\ket{\psi_n^r}\bra{\psi_n^l}}{\omega-\Omega_n},
\end{equation}
where $\ket{\psi_n^r}$ and $\bra{\psi_n^l}$ are right and left eigenvectors of the system's Hamiltonian, respectively and they satisfy the biorthogonality relation $\braket{\psi_n^l|\psi_m^r}=\delta_{nm}$.
In this case, in contrast to the Hermitian case, there are two points that should be noted: first, for exciting a mode (say $\ket{\psi_i^r}$), the input signal should have a non-zero overlap with the left eigenvector $\bra{\psi_n^l}$ (not with the Hermitian conjugate of the right eigenvector $\bra{\psi_i^r}$). Second, considering that generally the eigenvalues of the non-Hermitian systems are complex numbers, exciting the system with a frequency excitation matching the real part of one of the system's eigenfrequencies (say $\text{Re}(\Omega_i)$) does not cause the term proportional to $1/(\omega-\Omega_i)$ in the Green's operator expansion to diverge or become significantly dominant compared to the other terms. Consequently, due to the non-zero imaginary part of the eigenfrequency, matching the excitation frequency to the real part of the eigenfrequency does not necessarily ensure that the excitation signal is predominantly transferred to the desired mode ($\ket{\psi_i^r}$).
In a non-Hermitian system some modes may have a short lifetime, and as a result they will not appear as sharp resonances.

\subsection{\textit{Dependence on the edge termination}}
\begin{figure}
    \begin{center}
        \includegraphics[width=4.5in]{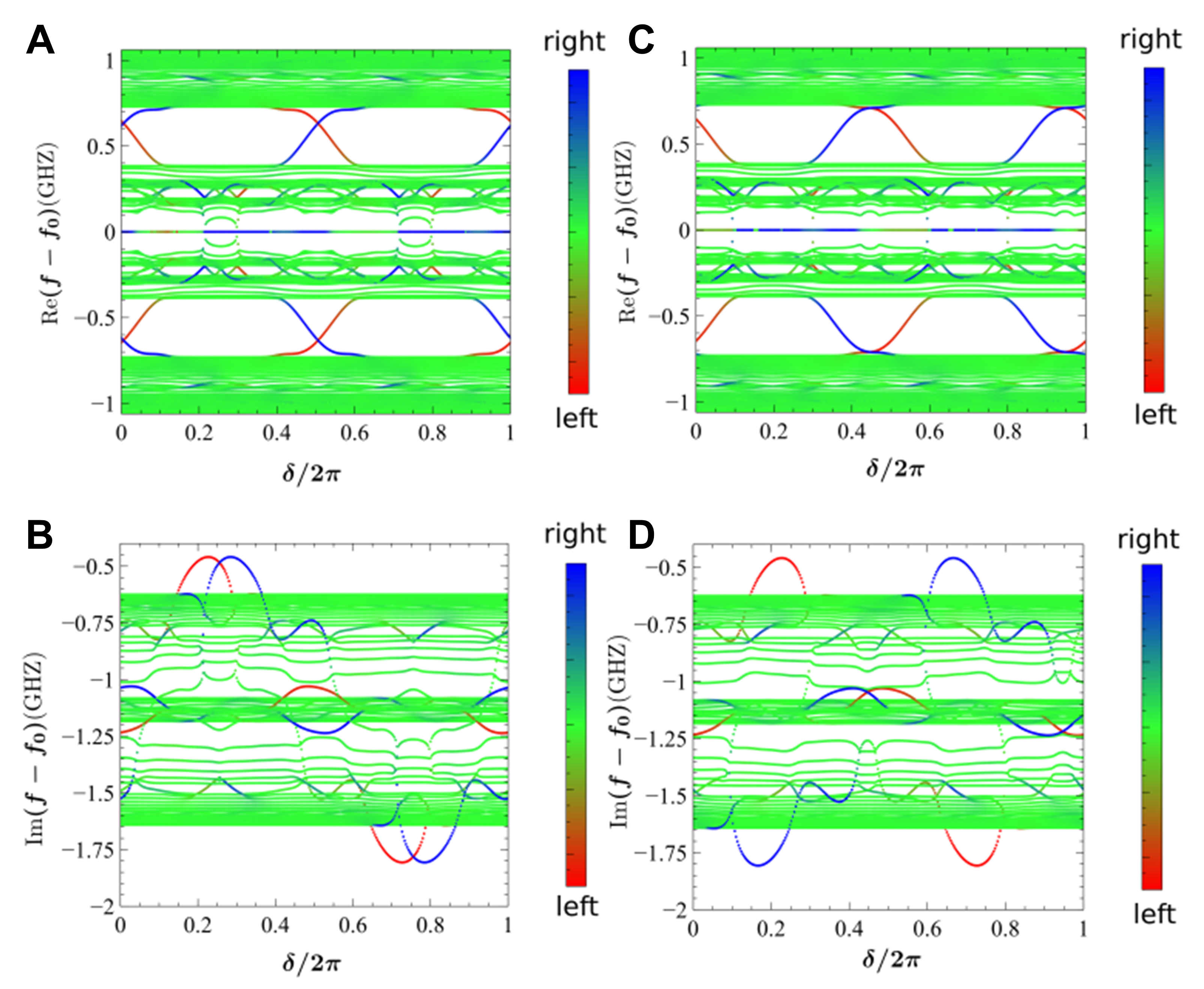}
        \caption{\textbf{The energy spectra of the incommensurate configuration. A, B}, show the real and imaginary part of the model spectra with a system size of $N=200$ sites, while {\bf{C}}, {\bf{D}}, show the real and imaginary spectra of the model with a system size of $N=201$ sites. The colors red, green, and blue indicate that the eigenvectors are localized at the left edge, in the bulk, and at the right edge of the non-Hermitian quasicrystal. The edge modes are sensitive to the system size, as it controls the termination,
        and the phase $\delta$ of the modulation potential.}
        \label{panel1}
    \end{center}
\end{figure} 
It is evident from the main text that, for the incommensurate case of the model, only one edge mode exists with $\text{Re}(E) = 0$. This edge mode was localized at the left edge of the system with $N = 8$. In the thermodynamic limit, a Hermitian quasi-periodic system has energy spectra with an infinite number of band gaps and an infinite number of edge modes. 
For generic modulation frequencies, the emergent edge modes do not appear in pairs
for a specific phase $\delta$, and they may appear at finite $\text{Re}(E)$. Moreover,
in small-sized systems such as the one in our experiment, the band gaps are sensitive to the boundary conditions, controlled by the length of the system and the
phase of the potential. In the thermodynamic limit,
edge modes remain sensitive to the specific potential value in the last site, as the local potential impacts the specific energy of the edge mode, as shown in Supplementary Fig. \ref{panel1}. 
In Supplementary Fig. \ref{panel1}, we plot the energy spectra for the system with two different values of $N$. The colors indicate the localization of the energy modes in the system: red and blue represent localization at the left and right edges respectively, while green indicates that the modes are delocalized throughout the chain. The two cases have different energy spectra, with edge modes localized at the left and right edges with different imaginary parts associated with them. The imaginary parts are responsible for the magnitude of the spectral density in the system,
which influences the experimentally measurable spectral function. 
It is also worth noting that edge modes coexist with bulk modes at zero energy. As a result,
for a specific system size, the system may feature a left edge mode and a bulk mode, both at $\text{Re}(E)=0$,
yet not a right mode. As the phase of the potential is changed, such a situation can be reversed.

\bibliography{Sup_reference}